\documentclass[sn-basic]{sn-jnl}

\usepackage{makecell}
\usepackage{tabularx}
\usepackage{eurosym}
\usepackage{wrapfig}
\usepackage{subfigure}
\usepackage{csquotes}

\usepackage{hyperref}
\usepackage{url}

\usepackage{float}

\newcommand{\CL}{\textit{CL}}
\newcommand{\owner}{\textit{owner}}

\newcommand{\damage}{\textit{Compensation}}

\newcommand{\arise}{\texttt{Arise}}
\newcommand{\duedate}{\texttt{DueDate}}
\newcommand{\limitation}{\texttt{Limitation}}
\newcommand{\Min}{\texttt{Min}}
\newcommand{\Max}{\texttt{Max}}

\renewcommand{\P}{{\cal P}}
\newcommand{\Os}{{\cal O}}
\newcommand{\Cs}{{\cal C}}
\newcommand{\SPA}{\textit{SPA}}

\definecolor{dkblue}{rgb}{0,0.0,1}
\definecolor{prismgreen}{rgb}{0, 0.6, 0} 
\lstdefinelanguage{pseudo}{ 
        basicstyle=\color{black}\small\ttfamily, 
        keywords= 
        {global, var, IF, ELSE,FOR,EACH,false,WHILE, true, function,ENDIF,THEN}, 
        keywordstyle={\small\bfseries\color{dkblue}}, 
        numberstyle=\small\ttfamily\bfseries\color{black}, 
        comment=[l] {//}, morecomment=[s]{/*}{*/}, 
        commentstyle= \color{prismgreen}, 
        tabsize=4, 
        captionpos=b, 
        xleftmargin=20pt,
        xrightmargin=5pt,
        frame=single, 
        escapechar=@ 
}

\jyear{2025}%

\theoremstyle{thmstyleone}%
\theoremstyle{thmstyletwo}%
\newtheorem{example}{Example}%

\theoremstyle{thmstylethree}%

\begin{document}

\title[Automated Consistency Analysis for Legal Contracts]{Automated Consistency Analysis
for\\ Legal Contracts}

\author[2]{\fnm{Alan} \sur{Khoja}}\email{Alan.Khoja@uni-konstanz.de}

\author[1]{\fnm{Martin} \sur{K\"olbl}}\email{Martin.Koelbl@uni-konstanz.de}

\author*[1]{\fnm{Stefan} \sur{Leue}}\email{Stefan.Leue@uni-konstanz.de}

\author[2]{\fnm{R\"udiger} \sur{Wilhelmi}}\email{Ruediger.Wilhelmi@uni-konstanz.de}

\affil*[1]{\orgdiv{Dept.\ of Computer Science}, \orgname{University of Konstanz}, 
\country{Germany}}

\affil[2]{\orgdiv{Dept.\ of Law}, \orgname{University of Konstanz}, 
\country{Germany}}

\abstract{
Business contracts, particularly sale and purchase agreements, often
contain a large number of clauses and are correspondingly long and complex.
In practice, it is therefore a great challenge to keep track of their legal
context and to identify and avoid inconsistencies in such contracts.
Against this background,
we describe a method and tool called 
\textit{ContractCheck}\footnote{\textit{ContractCheck} is available from \url{https://github.com/sen-uni-kn/ContractCheck}.} 
which allows
for the consistency analysis of legal contracts, in particular Share Purchase Agreements (SPAs). 
In order to identify the concepts that are relevant
for an analysis we define an ontology for SPAs.
The analysis is, then, based on an encoding of the preconditions 
for the execution of the 
clauses of an SPA, as well as on a set of proposed consistency constraints 
formalized using decidable fragments of First-Order Logic (FOL). 
Based on the ontology for SPAs, textual SPAs are first encoded in a structured natural language format that we refer to as ``blocks''. 
\textit{ContractCheck} interprets these blocks and constraints and
translates them into assertions formulated in FOL. It then invokes a
Satisfiability Modulo Theory (SMT) solver in order to check the 
executability of a considered contract, either by providing a satisfying model, 
or by proving the existence of conflicting 
clauses that prevent the contract from being executed. We illustrate 
the application of \textit{ContractCheck} to concrete SPAs,
including one example of an SPA of realistic size and complexity,
and conclude by suggesting
directions for future research.
}


\maketitle

\section{Introduction}\label{sec:intro}

\subsection{Motivation and Overview}
Contracts are essential in business. 
They allow the contracting parties to arrange their legal relationships 
by giving legal effect to their common will and establishing mutual claims. 
A prominent example is the purchase of a company 
in a share purchase agreement (SPA).
Like any contract of sale, an SPA must contain at least 
the indispensable \textit{essentialia negotii}: 
the purchaser and the seller, as well as 
the subject matter of the purchase and the purchase price to be paid.
In practice, SPAs regulate all relevant legal issues in the contract 
and exclude references to statutory law as far as possible.
As a result, SPAs have a very local semantics, based almost
entirely on the contractual obligations agreed in the SPA~\citep{OetkerMaultzsch.2018}.

Contracts, and especially SPAs, are often very long and complex. 
This is due to the large number and complexity
of the issues to be addressed. 
In addition, a large number of people are often involved in drafting contract texts.
Furthermore, the negotiation and the drafting process may extend over a long period of time 
and involve a large number of changes to the draft.
Length, complexity and frequent changes make a contract prone
to errors and inconsistencies, such as incorrect references,
missing essential elements and unfulfillable or unenforceable claims.
Inconsistencies in the form of missing essentials can be found 
by simple syntactic analysis.
It is much more complex
to find inconsistencies in the dynamics of
multiple claims.
Claims should not contradict each other and should be fulfillable and enforceable 
in the context of the legal facts described in the contract.
Also, the combination of several due dates may be
unexpectedly restricted by a statute of limitations stated in the contract.

For instance, assume that an SPA contains a warranty claim that
must be asserted within 14 days after the closing on day 28, 
then subsequent performance must be made within 28 days, 
and otherwise damages must be paid within another 14 days. 
Assume further 
that the contract contains a provision 
that warranty claims are limited to 42 days after closing. 
In this example, the warranty clause provides 
that performance may continue until day 84, 
while the limitation period ends on day 70.
This means that there are inconsistencies in the timing of the SPA.

We pursue a research agenda that aims at developing analysis methods
for the identification of logical inconsistencies, such as the ones
described above, in contract texts, with a specific focus on SPAs. It is our 
objective to develop automated methods and tools that do not require input
from the user, for instance guidance of mathematical proving or logical reasoning. We combine 
this objective with the goal to permit reasoning about a fairly wide range of
data domains, for instance natural or rational numbers, arithmetic expressions  
and uninterpreted functions representing uninterpreted logical predicates. 
This allows us to reason about, e.g., prices, dates,
timing, interest rates, compensation amounts and ownership properties. 
The identification of inconsistencies in contract texts is an objective per se. 
However, it also 
enables determining the executability of a contract in terms of sequences of
claim performances, or as a whole. 
As a use case for the developed method and tool we envision support for legal 
actors during contract drafting and negotiation, among others.

In this paper we propose a method and tool called
{\em ContractCheck}, designed to formally model and automatically
analyze a given SPA.
From a technical perspective, the {\em ContractCheck} approach is rooted in 
concepts from symbolic artificial intelligence,
such as symbolic logic reasoning and automated 
theorem proving, and 
inspired by experiences with the use of formal methods in the
analysis of software and system designs. 
We maintain that automated formal analysis 
enables the localization of erroneous or inconsistent contractual
text by automated logical reasoning, thereby significantly improving the quality of the contract document.
Experience shows that the adoption 
of formal analyses in practice is greatly enhanced by a) hiding the formality in the
analysis method, i.e., not confronting the end user with the need to express 
desired properties or the system model using a formal notation, and b) the use of fully
automated analysis methods that do not impose the burden of driving the analysis process 
by providing manual guidance on the user as it would, for instance, be required when using 
an interactive theorem prover. These insights have guided our development of the technical
aspects of \textit{ContractCheck}, in particular the automated form of logical reasoning 
that \textit{ContractCheck} relies on.

Against this backdrop, we propose to base the analysis of consistency properties in 
SPAs on a formalization of the claims that they encompass using decidable fragments of First-Order Logic (FOL), and to perform the analysis using Satisfiability Modulo Theory (SMT) 
solving techniques~\citep{KroStr16,DBLP:reference/mc/2018,5460875}. More precisely, we consider
an SPA to be a collection of claims. We define a weakest precondition style semantics 
for the satisfaction of claims that is based on enabling conditions for their 
executability, such as the timing constraints used in the above example. 
Based on this formalization, we define two types of logical consistency analysis questions,
namely \textit{Analyis I}, is every claim in the contract executable, and 
\textit{Analysis II}, is there a feasible execution of the SPA.
Answers to these questions rely on determining the satisfiability of the conjunction of the 
set of first-order formulae that formalize the claims and the contract execution.
The decision procedures 
required to analyze satisfiability of these formulae are efficiently 
implemented in various SMT solving tools, such as the Z3 solver~\citep{deMBjo08,5460875} developed 
at Microsoft Research, which we use in our analysis.

\subsection{Structure of the Paper}

After discussing some preliminaries in Section~\ref{sec:preliminaries},
we first develop an ontological meta-model for the relevant entities 
of an SPA and present this as a class
diagram from the Unified Modeling Language (UML)~\citep{UML}
in Section~\ref{sec:modeling}. 
Next, we define a semantics for the executability of 
clauses in an SPA using decidable fragments of 
quantifier-free FOL in Section~\ref{sec:formal}. 
The formalization of the various automated consistency
analyses performed by {\em ContractCheck} is 
presented in Section~\ref{sec:verify}.
Using the running example of the sale of a pretzel bakery, 
we then illustrate
the automatic consistency analysis of a given SPA 
using the SMT technology invoked
by {\em ContractCheck} in Section~\ref{sec:tool}.
We demonstrate the ability of \textit{ContractCheck} to detect and explain inconsistencies in a more complex, realistic 
SPA drawn from the literature in Section~\ref{sec:evaluation}.
We discuss threats to the validity of our results in Section~\ref{sec:threats}.
Finally, we conclude and outline future research in Section~\ref{sec:conclusion}.

\subsection{Scope and Methodological Considerations}
In this paper, we develop the conceptual basis for \textit{ContractCheck} using SPAs under German law, but this does not preclude adaptation to contracts with a different subject matter or under other jurisdictions, as they may be more complex but basically have a comparable structure with similar elements.

Looking at examples of SPAs, it is noticeable that they often consist of very similar blocks of text, differing only in certain parameters, such as the agreed price.
In the case of international company acquisitions, it is estimated that
about half of the text of the provisions is changed little or not at all~\citep{coates2016contracs}.
In order to achieve a formal logical representation of the SPA,
as part of our approach and tool we provide a library
of parameterized Structured English text blocks that can be freely
combined and used to compose the text of the contract. 
These blocks allow greater flexibility than conventional approaches, where contract creation is dialogue-driven and based on decision trees.
Each of these blocks has a formal semantics expressed by formulae from a decidable fragment of FOL.
The conjunction of these conditions then constitutes the logical
representation of the contract.
The analysis method that we describe in this paper generalizes to 
the class of all concrete contracts that can be formulated by 
composing and parameterizing the provided text blocks.

\subsection{Approach}

\begin{figure}[!ht]
  \centerline{
    \includegraphics[width=0.7\textwidth]{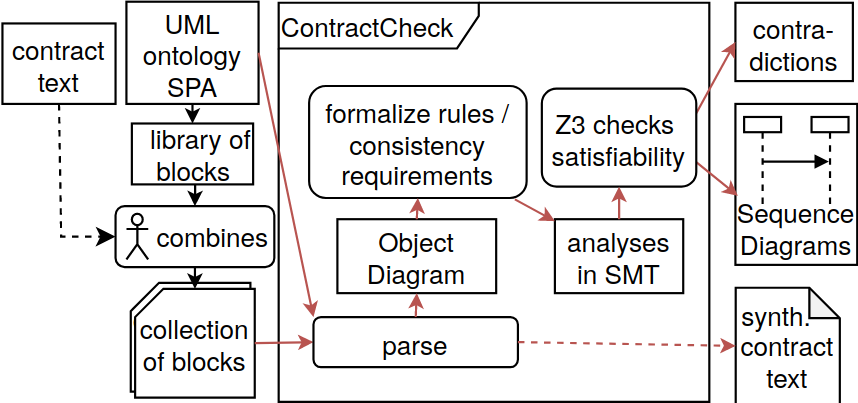}
  }
  \caption{The \textit{ContractCheck} approach towards automated contract analysis}
  \label{fig:ap_stm}
  \label{fig:approach}
\end{figure}

Figure~\ref{fig:ap_stm} sketches our proposed approach towards the
analysis of legal contracts.
First, we present a \texttt{UML Ontology} which provides a vocabulary
of entities from the domain of SPAs as well as 
a representation of their relationships.
Based on this ontology, the \textit{ContractCheck} tool provides a number of different
parameterized Structured English text skeletons, referred to as \texttt{blocks}, that can be used
to formally capture properties of the different clauses in an SPA.
From the perspective of contract generation, the idea underlying
this approach is that users are provided a collection of \texttt{blocks},
each representing a different type of contract clause, that they can then 
compile to form a textual contract document. The resulting collection of \texttt{blocks} 
can then be automatically be automatically analyzed by the \textit{ContractCheck} tool. Notice that in our experiments we do not (yet) take advantage of this contract generation feature. Instead, we are taking existing contract texts and translate them manually into collections of blocks for later analysis.

\textit{ContractCheck} implements an automated analysis workflow as indicated by the red arrows  
in Figure~\ref{fig:ap_stm}.
For analysis purposes, \textit{ContractCheck} parses the blocks,
determines the entities and relationships using the UML ontology and
generates an internal representation of the considered SPA in the form of an UML 
\texttt{Object Diagram}.
The syntactic correctness of the considered SPA is determined by checking the consistency of the object diagram
with the UML-defined ontology.
For dynamic consistency checks, several consistency \texttt{analyses} are
generated. In order to prepare the formal analysis, the object diagram is translated into 
\textit{SMT-LIB}~\citep{SMT-LIB} format, 
which the \texttt{Z3} SMT solver then checks for satisfiability, thereby determining logical consistency of the
clauses contained in the considered SPA.
The results obtained during the SMT solving 
are presented either by flagging the blocks that are determined to  contradict each other (Analysis I),
or by sequence diagrams representing possible SPA executions in case a contradiction-free execution of the 
SPA is possible (Analysis II).

\subsection{Related Work}
We review publications that address the logical modeling
and analysis of legal artifacts. 
We consider the 
verification of smart contracts, c.f.~\citep{Braegelmann.2019, Paal.2019}, 
which are effectively described by 
executable program code, to be outside the scope of this paper, since 
their analysis is more akin to program analysis.

\paragraph{Legal Modeling and Analysis}

Several meta-models of different legal domains aiming to describe legal entities and their relationships have been proposed.
A first approach towards representing contracts with UML has been proposed in~\cite{EngRitetal01}.
A multi-level hierarchical ontology 
relying on UML class diagrams in order to support the modeling of contracts is
presented in~\cite{KabJoh03}.
A modeling of contracts as business processes 
has been proposed in~\cite{Kab05, WanSin05}.
In~\cite{DesNarSin08}, a business process is translated into state
machines to check whether it is always beneficial for the contracting
parties to fulfill their claims during the execution of the contract.
Dynamic Condition Response Graphs~\citep{hildebrandt2011dcrgraphs} (DCR graphs) provides a graph-based 
language to model the dynamic dependencies and interactions between different legal conditions and 
responses along legal workflows. Temporal and logical properties of the event relations can be 
checked using formal analyses, including explicit-state model checking~\citep{DBLP:reference/mc/2018}.

\textit{LegalRuleML}~\citep{PalGovetal11, LegalRuleML, Grupp.2018} is a
specialization of the general language \textit{RuleML}~\citep{BolPasSha10}
for expressing relationships between legal entities by means of rules.
We are not aware of any extension of LegalRuleML to analyze contract executions for inconsistencies.
The Contract Specification Language (CSL) is
a modeling language that represents claims as actions in a contract. 
In~\cite{HviKlaZal12}, 
the execution of a given CSL expression is computed as a sequence of actions,
which are then analyzed to determine whether any specified obligations are met. 
In~\cite{HenLarMur20}, a contract is interpreted and analyzed as a
composition of commitments.   
In~\cite{MadKriKar14}, the natural language sentences of an
e-contract are translated into a dependency graph in order to check 
whether individual provisions contradict each other.

\textit{Deontic logic}~\citep{Wri51,sep-logic-deontic} is a family of modal logics 
designed to express the semantics of claims with
operators for the modalities of obligation,
permission, and prohibition. 
A Deontic Propositional Logic and its axiomatization are defined in~\cite{CasMai07}.
A tableau calculus for this logic is defined in~\cite{CasMai08}.
Tableau-based decision procedures for a class of logics 
encompassing temporal and deontic modalities are proposed in~\cite{BalBroBru09}.
These analyses find inconsistencies with respect to deontic or
temporal modalities within a contract, but do 
not calculate contract executions witnessing these inconsistencies.

The Contract Language $\CL$~\citep{PriSch07, PriSch12} encompasses 
deontic as well as dynamic propositional logic. 
In~\cite{PacPriSch07}, a contract given in $\CL$ is translated into
a transition system, which is then analyzed by the model checker NuSMV 
for inconsistencies. 
In~\cite{CamSch17}, a \textit{C-O diagram}~\citep{MarGreetal10}, 
which is a graphical extension of $\CL$, is expressed by
state machines and analyzed using the real-time model checker UPPAAL.
The analyses find syntactic inconsistencies and prove a 
dynamic inconsistency by a single contract execution.
The approach requires an expert to encode a contract as a \textit{C-O diagram}.
An extension of $\CL$ to $\textit{RCL}$ in~\cite{BonMur15} and \cite{BonMur21} proposes  
reasoning about the persons between whom claims exist
using the~\textit{RECALL} tool.

Another system supporting a deontic logic-based approach is the Linear Time 
Temporal Logic (LTL)-based language
\textit{FL}~\citep{GarMerSch10, GorMerSch11} together with the model 
checker \textit{FormaLex}~\citep{GorMerSch11, SchetAl17}.

\cite{schumann2024detection} review research on consistency analysis in regulatory
documents. They observe that the most frequently used class of techniques for detecting 
inconsistencies in the studies that they analyze are formal verification methods.
In this context, \cite{mitra2018identifying} describe consistency analysis of business
workflow rules based on an encoding of these rules in the SMT-LIB formalism and the 
performance of SMT solving on this encoding. Unlike our work, this
work does not analyze the consistency of legal contracts.

\paragraph{Programming and Domain Specific Languages}
\textit{Catala}~\citep{merigoux2021catala} is a functional programming language developed by Microsoft. It is specifically designed for use in a legal context and is intended to be used to automatically interpret and apply laws and regulations.
\textit{Flint/eFlint}~\citep{doesburg2016flint,binsbergen2020eflint} is a domain-specific language aiming at the formalization of normative legal texts. It is based on the framework
of legal fundamentals by Hohlfeld~\citep{Hohfeld}. The Flint/eFlint models can be executed and, to a 
limited extent, analyzed.
\textit{L4}~\citep{governatori2023l4}, also known as Governatori's domain-specific language for the legal domain, is a language that allows the expression of legal rules in a simple and structured way, using the syntax ``if precondition then conclusion''. It is intended to make it easier for legal experts to express legal rules in a way that can be understood by computers. 
\textit{Stipula}~\citep{sartor2021stipula} 
is a domain-specific language that allows for the creation of software legal contracts 
and smart legal contracts,
focusing on deontic concepts such as 
permissions, prohibitions, obligations, assets exchanges, escrows and securities.

\subsection{Comparison to Related Work}

The logical analysis approach that we propose in this paper focuses on the use
of decidable fragments of FOL in the formalization and the use of decision procedures for these fragments, implemented in 
SMT technology, to detect inconsistencies in SPAs.
While many of the approaches mentioned above do have the ability to 
detect inconsistencies in the execution of actions, they do not have the ability
to also produce 
models that reveal reasons for inconsistencies that lie in the data, which is an aspect 
that we focus on. 
For instance, in comparison to the cited approaches that use the UPPAAL real-time model 
checking tool, we are able to reason at the same time about inconsistencies in the timing of
the execution of claims and in the constraints on numerical values, such as
purchase prices and warranty claims.

Only finite executions are of interest in an SPA,
which means that LTL-style model checking capabilities of
infinite behaviors are not required for the purpose of our consistency 
analysis\footnote{Notice that we are not interested in,
for instance, proving that an SPA does not include non-progress cycles. Consequently, we do not
need to reason about liveness properties and can limit our analyses to safety properties, c.f.\ \cite{DBLP:journals/dc/AlpernS87}.}.
The types of analysis performed by \textit{FormaLex} are based on standard
LTL model checking and encompass a temporal interpretation of 
deontic modalities, such as obligation. The objectives of this work differ
from ours in that the analysis of domain data is not addressed in \textit{FormaLex},
while we are not focusing on reasoning about deontic modalities.

A further difference compared to the cited works lies in the ability of \textit{ContractCheck}
to prioritize primary over secondary claims, an important legal concept in SPAs, 
during Analysis II, which allows for a very meaningful type of executability analysis. 
The above cited approaches
are only able to logically link different claims without this type of prioritization.

\paragraph{Comparison to Legal Reasoning using Deontic Logics}
Reasoning based on deontic logics focuses on the modalities of obligation,
permission, and prohibition, which are ideally suited to reason about
properties of claims in legal contracts that need to be cast in terms
of these modalities. Reasoning in these logics means to 
derive inferences from the modalities obligatory, impermissible, optional, non-optional, permissible and obligatory in the "deontic hexagon"~\citep{sep-logic-deontic}. 
Depending on the choice of a syntactic fragment and the assumed semantics, deontic logics
may be prone to paradoxes, which limits the possibility of consistency checking, although numerous
fragments with decidable, consistent semantics have been defined, see for instance~\cite{PriSch12}.

While the logic of claims and the possible sequences of their performance,
which we refer to as contract executions, are a central concept in our paper, we 
are not interested in reasoning in terms of the above mentioned deontic modalities.
Our approach bears similarities with semantic reasoning about programs, where contract claims
play the role of program statements. We reason whether starting from an initial 
configuration of the contractual situation, due to the satisfaction of some logical condition 
(precondition), the performance of a claim is possible, and which logical result conditions 
(postcondition) its execution entails. The predicates that we use as pre- and postconditions
are formulated as FOL formulae over various decidable theories, such as 
integer and real arithmetic. This allows us to specify conditions on domain-specific values,
such as on due dates for claim performance, 
purchase prices or warranty compensations. We only choose fragments of FOL that are 
decidable and for which efficient decision procedures, implemented in SMT solvers such
as Z3, are available. These fully automated solvers are capable of returning models that provide 
evidence for why a contractual situation is inconsistent.

In conclusion, our approach differs from other analysis approaches relying on 
deontic logics in that it combines reasoning on possible contract executions, 
checking the consistency of domain-specific predicates, returns explanations for 
inconsistencies using domain-specific values, and is availing itself 
to efficient automated analysis through practical SMT-solving.

\subsection{Contributions}
This paper builds on results that appeared
in~\cite{DBLP:conf/spin/KhojaKLW22,khoja2022formal}. 
It presents an enhanced description of the method, a more comprehensive 
formalization and a more realistic case 
study.
In particular, the main contributions of the work documented in this paper are as follows:
\begin{enumerate}
    \item We propose an ontology for SPAs that we document using the UML.
    \item We provide a logical formalization of the semantics of SPAs using decidable fragments of 
    FOL. In doing so, we contribute to the extensive research area
    dealing with the use of formal logics in the representation of legal artefacts.
    \item We define a set of consistency analyses that are applicable to SPAs.
    \item We describe a tool that uses a collection of parameterized natural
    language building blocks from which textual SPAs can be composed. 
    The possible synthesis of contract text is not considered in this paper.
    \item We present the prototype tool called \textit{ContractCheck}, that uses SMT and satisfiability core
    technology to perform the consistency analyses and to produce
    diagnostic information explaining identified inconsistencies.
    \item We illustrate the application of this approach to an SPA of a complex example from legal practice to show that the method is effective and efficient.
\end{enumerate}

\section{Preliminaries}\label{sec:preliminaries}

We introduce key legal terms, the modeling language
used to describe the company sale contract, and the logic we use to formalize a contract.

\subsection{Basic legal terms}
Even if statutory provisions are largely excluded in an SPA, they are still relevant for the drafting and interpretation of the agreement \citep{Wilhelmi.2024a}.
This is because the contractual provisions are based on the underlying legal system and its systematics.
In particular, the  SPA that we use to illustrate our approach contains terms that can only be understood by reference to the 
legal provisions and concepts of the German Civil Code (BGB).

Against this backdrop, we explain some basic legal concepts of private law 
generally used in contracts, and in SPAs in particular. 
These concepts regard, first of all, the distinction between legal subjects and legal objects, 
the relationships between legal subjects and legal objects
and the possibility to change legal positions.

\textit{Legal subjects} are, on the one hand, \textit{natural persons}, i.e. human beings, 
and, on the other hand, \textit{legal persons}, in private law in particular associations 
and other corporate bodies such as the public or private limited company.
Only they can be the bearers of rights and obligations and, in particular, be the parties to a contract (of sale).

\textit{Legal objects} may be the subject of rights of control and use 
as well as rights of disposal of legal subjects.
In particular, they include \textit{things}, i.e., 
physical objects, but also \textit{rights}, such as \textit{claims} of one legal subject against another legal subject or company shares, and other assets, such as intangible assets.

\textit{Relations between legal subjects} are, in particular, \textit{claims}, 
i.e., the \textit{right} of one \textit{legal subject}
to demand an act or omission from another \textit{legal subject},
corresponding with the \textit{obligation} of the other \textit{legal subject}
to this act or omission.
In contracts, a distinction can be made between primary and secondary claims. 
\textit{Primary claims} are the original claims 
that exist even if the contract is performed without disturbance, 
while \textit{secondary claims} only arise if the performance of the contract is disturbed 
and primary claims or other promises are breached.
Such promises can be \textit{warranties} regarding the existence or non-existence of circumstances.
In case of a breach, they can provide a claim for subsequent \textit{performance}, 
\textit{restitution} or \textit{compensation}.

\textit{Relations between legal subjects and legal objects} are also possible. 
A paradigmatic example is the right of \textit{ownership} of an object. 
It basically gives the owner the right to deal with the thing at her/his discretion 
and to exclude others from exercising any influence whatsoever, 
and can be transferred to others \citep{Wilhelmi.2022c}.

It is also important to consider the possibility of \textit{changes in the legal position},
including the \textit{transfer}.
Claims can \textit{arise}, become \textit{due} or \textit{not enforceable}, change, or extinguish
as a result of actual events and legal transactions.
The ownership of an object can be \textit{transferred} from one legal subject to another, usually by means of a  contract.

In the case of an SPA, the object of purchase are the shares of a company (\textit{share deal}) or the assets held by a company (\textit{asset deal}) \citep{Wilhelmi.2024d}. The purchaser has a claim against the seller to transfer the ownership of the shares or the assets to her-/himself. Vice versa, the seller has a claim against the purchaser to transfer the purchase price. Both claims are primary claims and aimed at a transfer. In addition, a SPA usually contains \textit{warranty} claims as secondary claims that give the buyer certain rights against the seller if certain warranties are breached.

\subsection{Object Oriented Modeling}

We model the legal relationships of the entities in an SPA 
using an object-oriented modeling approach.
We illustrate this modeling using a fictitious SPA 
in which \texttt{Eva} sells the company \texttt{Bakery AG} to \texttt{Chris}.
In an object-oriented modeling approach, an object describes an element of the real world,
sometimes also referred to as an instance.
Instances of the same type are typically aggregated into classes, where the class 
then defines the type of the instances that form its membership. 
The Unified Modeling Language (UML)~\citep{UML} is the prevailing
modeling formalism for object oriented systems. 
For the models that we propose in this paper, \textit{class} and \textit{object} 
diagrams are the most relevant diagram types of the UML.
Class diagrams represent the classes to which objects are aggregated
as well as the interrelations among classes. Object diagrams depict
the relationships of object instances.

\begin{figure}[h!]
\subfigure[Class \texttt{SharePurchaseAgreement}]{
    \hspace{1cm}
    \includegraphics[width=2.8cm]{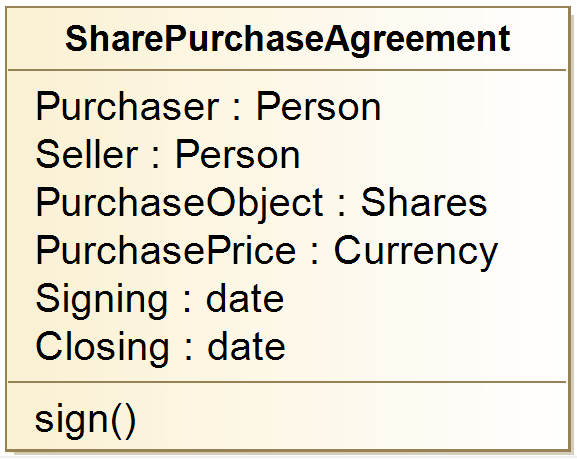}
\label{fig:stm_klasse}}
\hfill
\subfigure[Object \texttt{BakerySPA}]{
    \includegraphics[width=5.2cm]{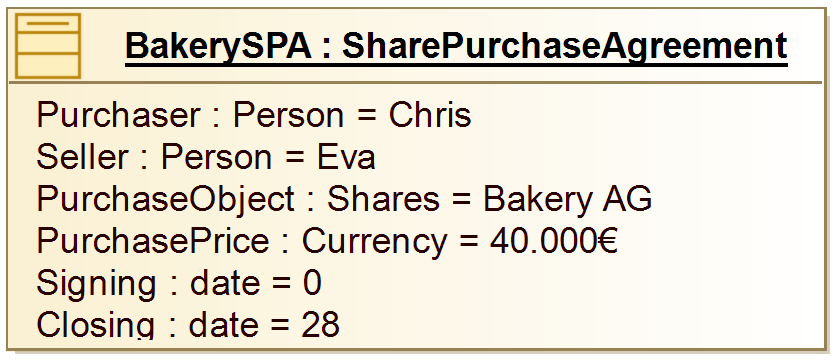}
\label{fig:stm_object}}
\caption{The object \texttt{BakerySPA} instantiates the class \texttt{SharePurchaseAgreement}.}
\label{fig:klasse}
\end{figure}

As illustrated in Figure~\ref{fig:klasse}, in the \texttt{Pretzel Bakery} example 
that we consider, the type of contract that we use is 
that of a \texttt{SharePurchaseAgreement}, 
which defines a class with that type, see 
the UML class diagram in Figure~\ref{fig:stm_klasse}.
A class can define \textit{attributes}
in the first box below the class name. 
An attribute represents properties of an object 
and has a name and a type.
For example, an SPA will have an attribute
\texttt{signing} of type \texttt{date}, which contains the date when the 
contract was signed, see Figure~\ref{fig:stm_klasse}.
The optional box beneath the attributes lists the
\textit{operations} that the elements of a class can perform. 
An operation represents a capability of the 
objects of some class, in the example in Figure~\ref{fig:stm_klasse} for
instance the ability that a contract can be signed, indicated by the operation
\texttt{sign()}.
It sets the attribute \texttt{signing} to the value $0$.
In the later analyses, it is assumed that the contract 
starts on day $0$ and this value will be assigned to a logical 
variable in the formalization of the SPA.

A UML object diagram 
illustrates possible values that the attributes of a particular instance
of some class can attain.
The object \texttt{BakerySPA} in Figure~\ref{fig:stm_object} describes an instance of the class \texttt{Share Purchase Agreement}
and represents the fictitious SPA that we use for illustration purposes in this paper, see Figure~\ref{fig:stm_object}. Notice that the object \texttt{BakerySPA} carries concrete values
for the various attributes defined in the class \texttt{Share Purchase Agreement}, e.g., the
names of the \texttt{Purchaser} and the \texttt{Seller}.

\begin{figure}[h!]
  \centering
  \includegraphics[width=0.85\textwidth]{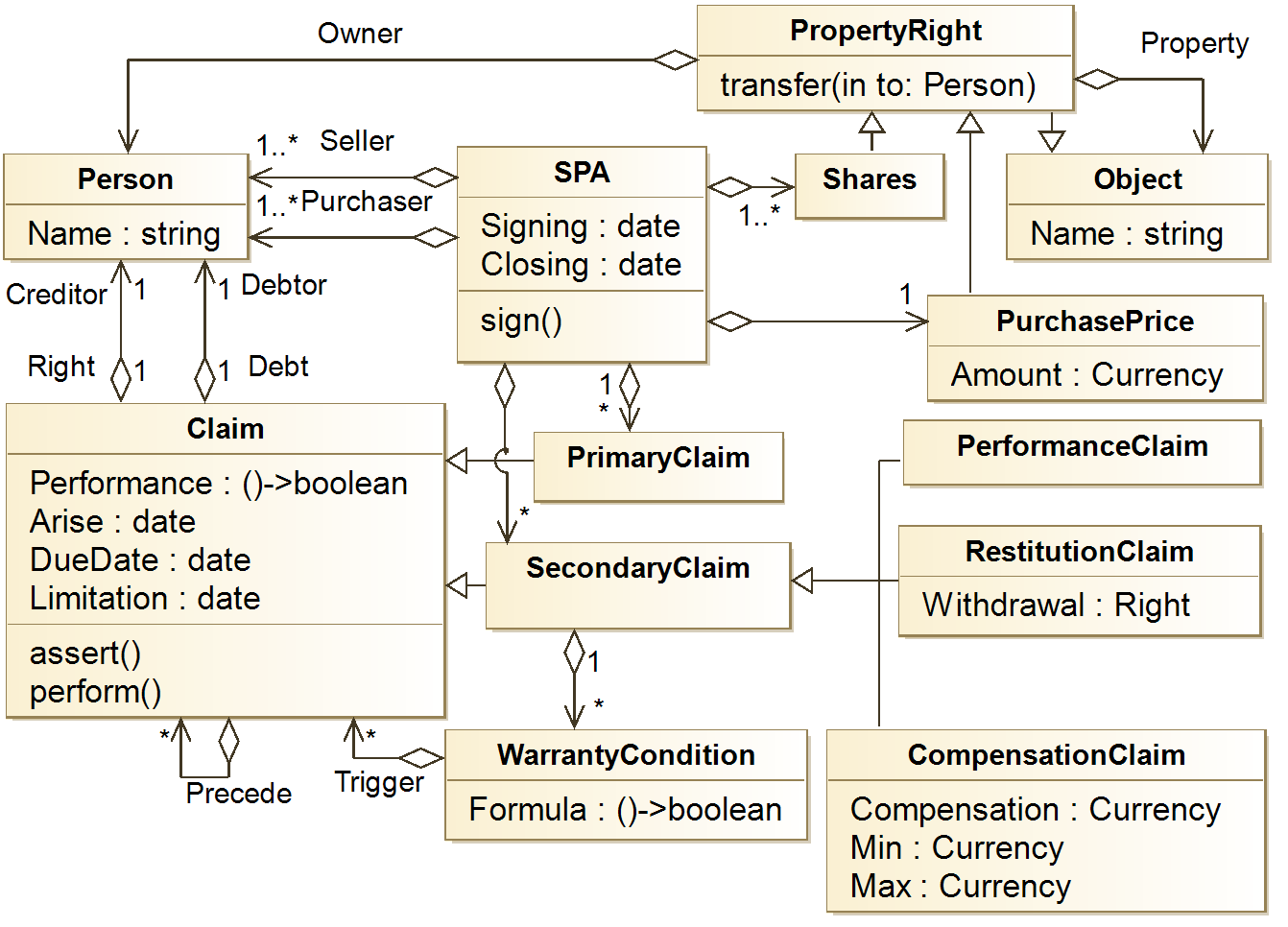}
  \caption{Class Diagram of a Share Purchase Agreement (SPA)}
  \label{fig:contract}
\end{figure}

Central to the notion of object-oriented modeling is the idea of modeling at different
levels of abstraction, using different views on a system.
In this context, an important technique is that of refinement, where a more abstract view of 
a system is replaced by a more concrete one, for instance one that reveals more detail of
the internal structure of the considered system. In particular, refinement 
can be achieved by replacing an abstract view on the system, for instance given by a 
single class, by a more concrete view defined by a collection of more concrete
classes as well as a description of the relationship among these more concrete classes.
A further type of refinement is accomplished in object-oriented modeling by the 
concept of specialization through inheritance. Here, a more abstract class is specialized
into a number of more concrete classes, so that the more concrete classes inherit the 
attributes and operations of the parent class, while additional attributes and operations
can be added to the more concrete class.

UML class diagrams permit the modeling of systems using refinement, specialization
and aggregation. Classes are represented by rectangles, and relationships among classes,
referred to as associations, are depicted by lines connecting classes. These lines 
may have text labels denoting the name of the association, the role of the classes as well 
as the cardinalities of the association at either end of the association. Further
triangle- and diamond-shaped labels qualify an association as representing inheritance and
aggregation, respectively. In the following we will illustrate these concepts in principle 
using the class diagram in Figure~\ref{fig:contract}, the meaning of which will be explained in more 
detail in Section~\ref{sec:modeling}. The class diagram in Figure~\ref{fig:contract} represents a refined, more concrete view 
at the concept of a Share Purchase Agreement. Central is the concept of a claim, that we use
here to illustrate the above mentioned UML modeling concepts.

    The class \texttt{Claim} is \textit{specialized} through \textit{inheritance} into either a class \texttt{PrimaryClaim}, or a 
    class \texttt{Secondary Claim}, as indicated by the association lines that connect these
    classes and that begin with a triangle on the side of the more general class, pointing to it. Inheritance means
    that all attributes and operations of the inheriting class are inherited by the specialized 
    class. For instance, the class \texttt{PrimaryClaim} will inherit attribute \texttt{DueDate} as
    well as operation \texttt{assert} from class \texttt{Claim}. Notice that the reverse direction of 
    \textit{specialization} is also referred to as \textit{generalization}, i.e., the classes 
    \texttt{PrimaryClaim} and \texttt{Secondary Claim} can be generalized to the class \texttt{Claim}.
 
 The association type \textit{aggregation} represents a part-whole relationship. As 
    indicated by the aggregation associations between classes \texttt{Claim} and \texttt{Person}, a claim
    consists of one reference to a person where the person has the role \texttt{Creditor}, and another
    where the person has the role \texttt{Debtor}. In the first case the claim is a \texttt{Right} and
    in the second case a \texttt{Debt}. The cardinalities of the associations are indicated by the 
    numbers or ranges at either end of an association. For instance, there is a 1:1 relation between
    \texttt{Claim} and \texttt{Person} along the \texttt{Right} / \texttt{Creditor} aggregation association.
    
\subsection{Decidable Fragments of First-Order Logic}

We formalize the logical constraints implied by the 
clauses of an SPA in decidable fragments of FOL.
A fragment of FOL is decidable if 
a complete and terminating algorithm exists that decides for any formula from this fragment,
whether the formula is satisfiable.
General FOL is undecidable,
which is why we restrict our modeling to decidable fragments of it.

The formalization in this paper uses 
an FOL fragment which includes 
the decidable theories of linear real arithmetic, 
equality, integers and uninterpreted functions~\citep{KroStr16}.
Efficient SMT solvers~\citep{KroStr16}, which automatically
decide the satisfiability of formulae from these fragments, are available.
In case of satisfiability, they return a satisfying assignment, also referred to as a model, of values to the
logical variables.
Otherwise, they return the result {\em unsat} as well 
as an unsatisfiability core, which contains an explanation
for the unsatisfiability.
For instance, an SMT solver determines that the formula
$(x>0) \wedge (x+y<0)$ is satisfiable and may, for example, return the variable 
assignment $x=0.9$ and $y=-2.0$ as a satisfying model.
When the considered logical constraint system is structured as a partial MaxSMT instance
using what is referred to as ``soft assertions",
some SMT solvers, such as for instance Z3,
have the ability to return a maximal set of constraints
that are necessary to satisfy an otherwise unsatisfiable
model. We will take advantage of this 
capability when prioritizing a certain type of claim in the analysis.
\section{Contract Modeling}\label{sec:modeling}

We propose the use of UML class diagrams in the ontological modeling of an abstract view of an SPA.
An instance of a class diagram is then used 
to represent a concrete legal contract. 
For legal entities that an expert finds in a concrete legal contract, s/he instantiates classes and adds values and relations given in the contract by assigning attributes.
The combination of the objects and their assignments is represented using an object diagram.
This information will be inserted in the blocks which are 
then automatically processed by the \textit{ContractCheck}
tool.
The set of blocks obtained in this way will be the basis for the logical consistency analyses 
that we present in later sections.
We illustrate this approach by applying it to the modeling 
of a concrete example SPA regulating the purchase of a pretzel bakery.

\subsection{Modeling of an SPA}

For modeling purposes, it is necessary to determine the typical
provisions in an SPA. 
These can be found in numerous legal template books.
For the present project, we have analysed a large number of templates in 
a number of 
form book on German law~\citep{Hoyenberg.2020,
Hoyenberg.2020b, MeyerSparenberg.2022a, MeyerSparenberg.2022b,
Pfisterer.2022, Seibt.2018a, Seibt.2018b, Seibt.2018c, Seibt.2018d}.
We have identified the following provisions as typical for an SPA: 
contracting parties, subject matter of purchase, purchase price provisions, 
conditions and execution, warranties and indemnities, liability, 
and final provisions.

Based on this, an example SPA containing these typical provisions,
which governs the sale of a pretzel bakery, 
has been developed.
We use this SPA as a running example in
this paper.
We are mainly concerned with the provisions related to \textit{warranties} and \textit{liability}\footnote{
We set legal terms in {\em italics}, and terms referring to UML diagrams in {\texttt teletype} font.}, 
supplemented by the indispensable provisions concerning 
the contracting parties (\textit{purchaser} and \textit{seller}), 
the \textit{purchase object} and the \textit{purchase price}.

From these provisions, we derive an ontology of the provisions in an SPA 
that is given as a UML 
class diagram, presented in Figure~\ref{fig:contract}. 
It represents the provisions of an SPA in the form of UML classes.
Note that this ontology can be extended to other types of contracts.
In this paper, we restrict ourselves to considering SPAs since, 
compared to other types of contracts, they
rely much less on implicit legal facts implied by legal dogmatics,
since these are based on the statutory provisions,
which are typically abandoned in SPAs.

\begin{figure}[htb]
\centerline{
\includegraphics[width=0.9\textwidth]{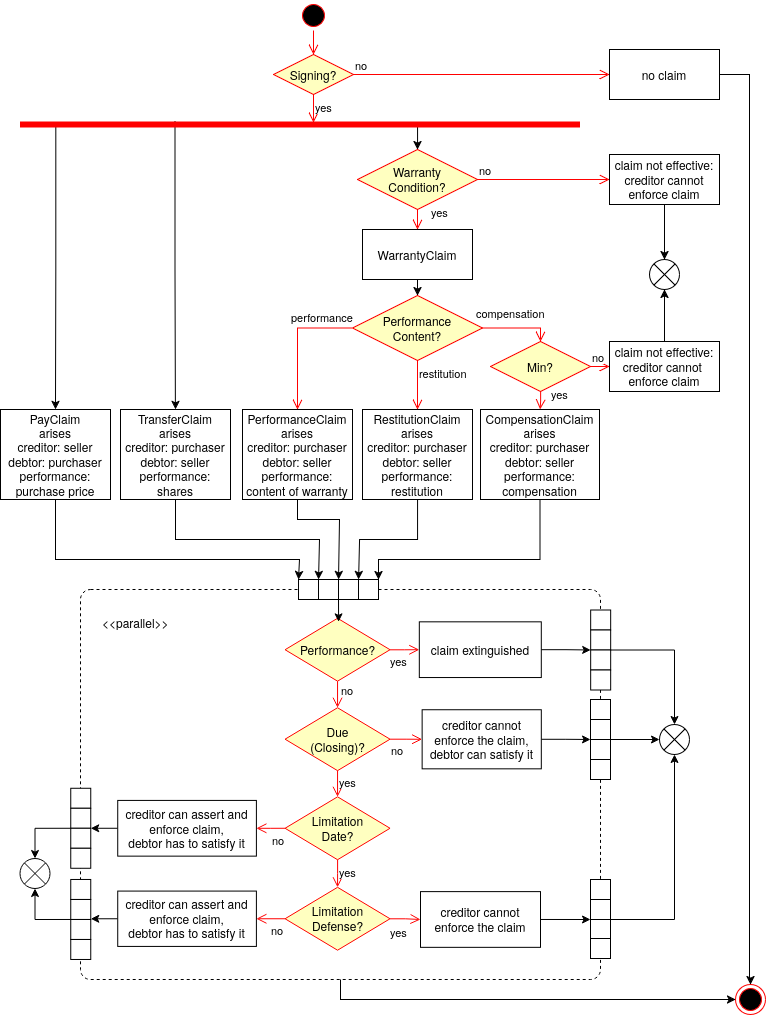}
}
\caption{Claim arise and fulfillment in the Bakery Contract.}
\label{fig:spa_activity}
\end{figure}

The essential \textit{claims} in an SPA are 
the \textit{claims} between the \textit{purchaser} and \textit{seller} 
for the \textit{shares} in the company and the \textit{purchase price} 
as well as related \textit{claims} from \textit{warranties}. 
We illustrate their relationship using a UML Activity
Diagram depicted in Figure~\ref{fig:spa_activity}.

When the \textit{contract} is \textit{signed} (\texttt{Signing}), 
the \textit{claim} for payment of the \textit{purchase price} (\texttt{PayClaim})
and the \textit{claim} for transfer of the \textit{shares} (\texttt{TransferClaim}) \textit{arise}.
In contrast, a \textit{claim} resulting from a warranty (\texttt{WarrantyClaim}) 
only becomes effective if additionally a condition is satisfied.

For each \texttt{Claim} one person (\texttt{Creditor}) can demand the \textit{performance} 
and another person (\texttt{Debtor}) owes the \textit{performance} as a duty.
The \textit{performance} can be a behavior or the production of an outcome (\texttt{Performance}).
For the \texttt{PayClaim}, the \texttt{Creditor} is the \textit{seller}, 
the \texttt{Debtor} is the \textit{purchaser}
and the \texttt{Performance} is the payment of the \textit{purchase price} 
by the \textit{purchaser} to the \textit{seller}.
For the \texttt{TransferClaim}, the \texttt{Creditor} is the \textit{purchaser}, 
the \texttt{Debtor} is the \textit{seller}
and the \texttt{Performance} is the transfer of the \textit{shares} 
by the \textit{seller} to the \textit{purchaser}. 

A \texttt{WarrantyClaim} additionally requires the breach of the \textit{warranty} 
as \texttt{WarrantyCondition} to become effective.
Otherwise, the \textit{creditor} cannot enforce it and its assessment ends,
whereas the assessment of the other \texttt{Claims} is independent of this.
\texttt{WarrantyClaims} can be distinguished into three types, 
depending on the content of their \textit{performance} (\texttt{PerformanceContent}):
the \textit{performance} of the content of the \textit{warranty} (\texttt{PerformanceClaim}), 
the \textit{restitution} of the \textit{purchase price} (\texttt{RestitutionClaim}) and 
the \textit{compensation} of the damages of the \textit{purchaser} (\texttt{CompensationClaim}).
A \texttt{CompensationClaim} is often additionally limited in the way
that the damage must reach a minimum value (\texttt{Min}),
otherwise the \textit{creditor} cannot enforce it and its assessment ends,
whereas the assessment of the other \texttt{Claims} is independent of this.
For the the three \texttt{WarrantyClaims}, 
the \texttt{creditor} is the \textit{purchaser} and 
the \texttt{debtor} is the \textit{seller}.

Each of these \texttt{Claims} independently not only needs to have \textit{arisen}, 
but also needs to have not \textit{extinguished}, be \textit{due} 
and not yet fall under the \textit{limitation}.
If the owed \texttt{Performance} is rendered, the \texttt{Claim} extinguishes 
and the \texttt{creditor} cannot assert (and enforce) it.
If the \texttt{Claim} is not yet \texttt{Due}, which here means before the \texttt{Closing},
the \texttt{creditor} cannot yet enforce the \texttt{Claim}, 
but the \texttt{debtor} can still satisfy it in rendering the \texttt{performance}.
If the \texttt{Claim} is due, the \texttt{creditor} can assert and enforce it 
and the \texttt{debtor} has to satisfy it up to the \texttt{LimitationDate}.
From the \texttt{LimitationDate}, the \texttt{creditor} can enforce the \texttt{Claim} only,
if the \texttt{debtor} does not raise the \textit{defense of the statute of limitations} (\texttt{LimitationDefense}), 
otherwise the \texttt{creditor} cannot enforce the \texttt{Claim}.
The respective run ends when all \texttt{Claims} have been examined. 
The \textit{contract} ends when all \texttt{Claims}
have extinguished or are no longer enforceable.

In the UML class diagram, depicted in Figure~\ref{fig:contract}, 
each \texttt{SPA} has at least one \texttt{Person} as \texttt{Seller} 
and another \texttt{Person} as \texttt{Buyer}.
For each \texttt{Claim}, either the \texttt{Purchaser} or the \texttt{Seller}
is the \texttt{Debtor}, who owes the \texttt{Performance} as \texttt{Debt}, 
and the other one is the \texttt{Creditor}, who has the \texttt{Right} to demand the \texttt{Performance} 
in \texttt{asserting} the \texttt{Claim}. 
For each \texttt{Claim}, either the \texttt{Purchaser} or the \texttt{Seller}
is the \texttt{Debtor} and the other one is the \texttt{Creditor}.
The \texttt{Creditor} has the \texttt{Right} to demand, 
that the \texttt{Debtor} pays his \texttt{Debt} in rendering the \texttt{Performance}.
The exercise of this \texttt{Right} is described as the operation \texttt{assert}.
The promised \texttt{Performance} is described as an attribute in the \texttt{Claim}.
The date from which the \texttt{Claim} is due, 
i.e. the date from which the \texttt{Creditor} can \texttt{assert} it, 
is described with the attribute \texttt{DueDate}.
The date from which the \texttt{Debtor} is entitled 
to refuse \texttt{Performance} of the \texttt{Claim}
in raising the \textit{defense of the statute of limitations}
is described with the attribute \texttt{LimitationDate}.
A \textit{claim} is extinguished if the \textit{performance} owed is rendered. 

We distinguish between \texttt{PrimaryClaims} and \texttt{SecondaryClaims}.
\texttt{PrimaryClaims} contain the \texttt{Performances} that are typical for the contract. 
They are performed if there is no disturbance of the contract. 
The \textit{claims} to transfer the \textit{shares} and the \textit{purchase price} 
are \texttt{PrimaryClaim}s. 
To cover their \texttt{Performance}, we need to model 
the \texttt{Shares} and \texttt{PurchasePrice} 
and their generalization \texttt{property right},
because the \texttt{Debtor} \texttt{performs} the \texttt{Claim}, when s/he is
the \texttt{Owner} of the \texttt{PropertyRight} 
in the \texttt{Shares} or the \texttt{PurchasePrice} and 
\texttt{transfers} this right to the \texttt{Creditor}.
The \texttt{PrimaryClaim}s have to be performed from the outset by the \texttt{DueDate},
sometimes under the condition of the \texttt{Performance} of another \texttt{PrimaryClaim}.
In contrast, \texttt{SecondaryClaims} arise only if the contract is disturbed.
They are subject to a specific condition, which we refer to as a \texttt{WarrantyCondition}.
This condition may consist of the breach of another \textit{claim}, often a \texttt{PrimaryClaim}.
It may also consist of the breach of separately enumerated circumstances 
as a condition independent of the breach of another \textit{claim}, 
which is typical of SPAs.
We model these conditions with the attribute \texttt{Formula}.
\texttt{SecondaryClaims} for which the \texttt{WarrantyCondition} entails the breach of another \texttt{Claim},
are \textit{consequence claims}.
In this case, the link between the \texttt{WarrantyCondition} and the other \texttt{Claim}
is represented by an association named \texttt{Trigger}.
\texttt{SecondaryClaims} for which the \texttt{WarrantyCondition} only entails separately enumerated circumstances,
are called \textit{independent warranty claims}. 
They are \textit{independent claims} if they do not refer to other claims.

The most important \textit{secondary claims} in an SPA 
are \textit{warranty claims}~\citep{Wilhelmi.2024b}.
A \texttt{WarrantyClaim} relates to an unknown risk arising from a breach of a
\textit{primary claim}, or from listed circumstances 
that are not expected but are considered probable enough to require regulation.
A \textit{warranty} consists of the \textit{warranty content} as a prerequisite and 
one or more \textit{warranty claims} as a consequence of a breach of the \textit{warranty}.
The \textit{warranty content} includes the existence or non-existence of certain circumstances, 
which usually concern the \textit{purchase object} and, in particular, its properties. 
It contains the \texttt{WarrantyCondition} for the warranty claims.
They are usually \textit{claims} of the \textit{purchaser}
for monetary compensation for the damage caused by the breach of the \texttt{WarrantyCondition}, 
irrespective of fault (\texttt{CompensationClaim}).
It is also possible to agree on a \textit{claim} for subsequent performance
(\texttt{PerformanceClaim}) or a right of \texttt{Withdrawal}
which, in the case of a notice of withdrawal (\texttt{noticeWithdraw}),
terminates the contract and may give the purchaser a \textit{claim}
for restitution of the \textit{purchase price} 
and the seller a \textit{claim} for restitution of the \textit{shares} (\texttt{RestitutionClaim}s).
A \texttt{CompensationClaim} is often limited in the way 
that it only arises, if the \textit{Damage} reaches a minimum value \texttt{Min}, 
and the \texttt{Compensation} is capped at a maximum value \texttt{Max}.
The limitation may apply only to individual \textit{claims} 
or to all \textit{claims} under the contract.
If the \textit{purchaser} has a \textit{claim} for compensation, 
this may be set off against the purchase price claim, 
so that the total amount of money to be paid is reduced accordingly.

\begin{figure}[htb]
  \centering
  \includegraphics[width=0.99\textwidth]{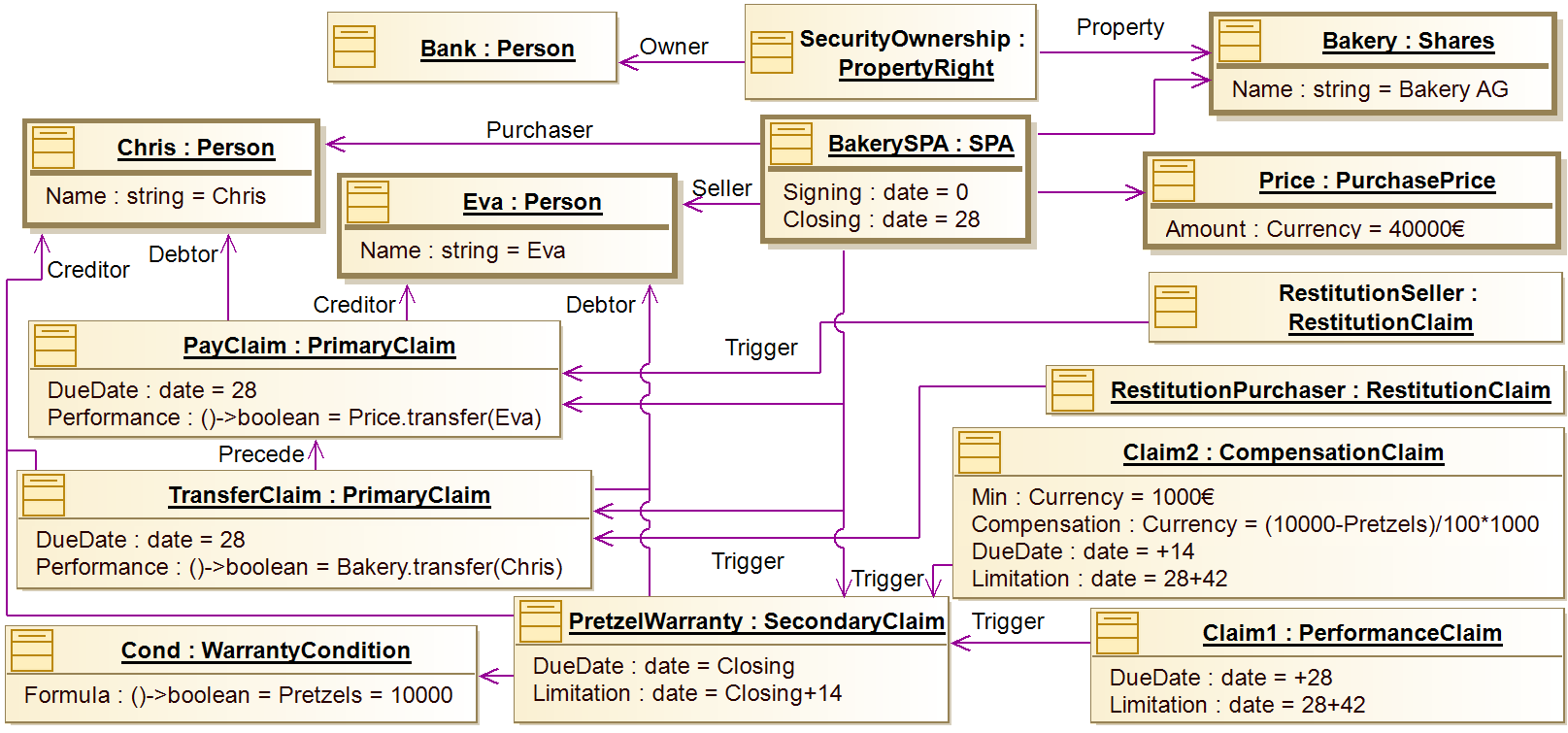}
  \caption{Object Diagram of Bakery SPA (with only partial depiction of the associations to avoid clutter)}
  \label{fig:example}
\end{figure}

The due date of a consequence claim is usually relative to 
the assertion date of a warranty.
The contract model supports absolute and relative time relations.
Relative time differences start with the plus sign $+$ and are relative
towards the occurring or asserting date of the claim to which the \texttt{Trigger} refers.
Examples are given in Figure~\ref{fig:example}. The \texttt{PerformanceClaim}
has to be performed within $+28$ days after the \texttt{PretzelWarranty} was asserted.
The limitation $28+42$ of the claim is given as
an absolute value.

A claim can also have an association \texttt{Precede} to another claim.
A \textit{claim} does not become due without its \textit{preceding claim} being performed beforehand.
For instance, the payment of the \textit{purchase price} may be expected to occur 
before the transfer of the \textit{shares}.
Otherwise, the seller can defense the share transfer because of the lack of payment.
In this case, the \texttt{Claim} for the \texttt{PurchasePrice} is a \textit{preceding claim} 
that has the \textit{performance} of the \texttt{Claim} for the \texttt{Shares} as a condition.
In the diagram in Figure~\ref{fig:contract}, the link between the \textit{claim} 
and the \textit{preceding claim} on whose occurrence it defenses
is represented by a self-association named \texttt{Precede}.

\begin{example}[Bakery SPA]\label{exa:pretzel}
As a running example, we use an SPA for the fictitious sale of
a pretzel bakery from the seller \texttt{Eva} to the purchaser \texttt{Chris}\footnote{The text of the Bakery SPA is given in the Appendix.}.
This SPA is a concrete instance of the SPA type depicted in the class diagram
in Figure~\ref{fig:contract}. This concrete instance is shown in the UML
class diagram 
presented in Figure~\ref{fig:example}.
In the bakery SPA, Chris agrees to pay the purchase price of \euro $40,000$  (\texttt{PayClaim}), 
and Eva agrees to give Chris ownership of the bakery 
by transferring \texttt{Shares} of the \texttt{Bakery}, which
constitutes the \texttt{TransferClaim}.
Both claims are due on the agreed closing date of $28$ days after signing.
If the pretzel bakery is not transferred or the purchase price 
is not paid at closing, the other party may 
withdraw from the purchase agreement.
The conditions about arise and fulfillment of claim in the Bakery SPA is depicted in the activity diagram in Figure~\ref{fig:spa_activity}.

In addition to the \textit{primary claims}, 
Eva provides a warranty in a \texttt{WarrantyClaim} named \texttt{PretzelWarranty}
which states that the bakery can bake at least $10,000$ pretzels a day.
In the example, if the bakery cannot bake at least $10,000$  pretzels a day, 
then the warranty is breached and Chris must assert this breach 
within the \texttt{DueDate} of $+14$ days after closing.
In the event of an assertion, Eva has to make good within $+28$ days
because of the \texttt{PerformanceClaim} \texttt{Claim1}, 
otherwise she has to pay within $+14$ days a compensation of \euro $1,000$
per $100$ of pretzels that cannot be baked, due to the \texttt{CompensationClaim}
\texttt{Claim2}, which entails a minimal compensation of \euro $1000$.
Any claim under the warranty has a \texttt{Limitation} of $42$ days from closing. 
Other crucial facts are represented in the bakery SPA.
Because of Eva's debts, a local bank \texttt{Bank} 
has a security interest in the shares of the bakery \texttt{SecurityOwnership}. 
Under German law, this means that the bank is the (security) owner of the bakery.

Manual inspection reveals the following inconsistencies in the SPA:
Eva's \textit{primary claim} according to the SPA is to transfer
the bakery to Chris.
To satisfy her claim, she must be the owner.
Note, however, that the Bank is the owner of the bakery
by virtue of the transfer of ownership by way of security.
Eva cannot fulfill her claim because she is no longer the owner of the bakery, which we consider an inconsistency.
While there are legal ways for dealing with this situation, the objective of our 
consistency analysis will be to point out such stipulations that require further attention.
Another inconsistency is due to the timing of the claims.
The \texttt{PretzelWarranty} has a \texttt{Limitation} of $42$ days but if an
assertion occurs within $14$ days after closing,
the consequence \texttt{Claim1} takes up to $28$ days and \texttt{Claim2} another $14$ days. This implies that \texttt{Claim2} can take up to $56$ days after closing,
contrary to the \texttt{Limitation} of $42$ days.
\end{example}

\section{Formalization}\label{sec:formal}

We present the logical encoding of the objects of an SPA in decidable 
fragments of FOL, as this turns out to be ideally suited to 
accommodate consistency checking for a large variety of possible legal 
scenarios, including ones that are not known at the time of signing. 
The versatility of available decidable first-order theories 
allows us to capture a large range of domain constraints in SPAs.

We then demonstrate the applicability of the proposed encoding
using the pretzel bakery SPA case study that we introduced above. 
The aim of this encoding is to enable a model-based satisfiability analysis
to determine logical consistency of the provisions of an SPA. 
A satisfying model for  the 
logical encoding represents a possible execution of the 
considered SPA. 
An unsatisfiable encoding, on the other hand, hints at either inconsistent claims
or even a non-executable contract.
We intend the consistency analyses in Section~\ref{sec:verify} to be performed primarily before signing, 
when the legal facts are only partially known. 

\subsection{Logical Formalization of Contract Entities}

The quintessential concept of a contract is that of mutual claims, i.e., the right of a 
legal subject to demand an act or omission from another legal subject. 
An SPA consists of a set of claims with which an execution
of the SPA must comply.
Whether a claim is performed in an execution depends on the objective situation and the behavior of the contracting parties, which we call \textit{legal facts}.
We use variables over various domains to model these facts, such as 
integers to model purchase prices and dates, and first order predicates to model property relations.
We restrict
their possible values according to the constraints specified in the SPA. 
These facts are then used to specify pre- and postconditions for the execution of claims
as they can be extracted from the SPA.

We define the \textit{execution} of a contract as the performance of a sequence of claims.
A particular execution of a claim is determined by a combination of claims and
the satisfaction of their preconditions, which includes factual and legal information 
given by the provisions of the contract,
such that for each \textit{primary claim} or \textit{independent claim},
the claim itself or one of its associated \textit{consequence claims}
is performed. 
The pre- and postconditions for the execution of the claims define a 
weakest precondition semantics of the contract stipulations
in the sense that the logical formulae that use in the formalization capture a precondition for the executability of a claim, for instance that 
the due date of a particular claim has been reached, and a postcondition that expresses that this claim has now been
executed.
Note that not every claim in an SPA must be performed
in a particular given execution.

The types of inconsistencies in claims and among claims
encompass the following:
\begin{itemize}
    \item primary and secondary claims may be performed outside
    the time interval between their due dates and the limitation
    date specified in the contract,
    \item a secondary claim is not performed in this time 
    interval in spite of the fact that the corresponding
    primary claim is not performed,
    \item the compensation performed by executing a warranty claim
    is not within the limits specified in the contract,
    \item numerical values, such as the purchase price
    or a numerical performance characteristic, are
    not within specified limits, and
    \item an ownership relation is violated, e.g., by 
    selling an item that the seller is not the owner of.
\end{itemize}
Notice that this list is not claiming to exhaustively 
cover all possible inconsistencies in contract texts.
Rather, it covers a large portion of inconsistencies that
are encountered in legal practice.

First, in Section~\ref{sec:facts} we formalize the legal facts, then in Section~\ref{sec:form_claim} we provide 
abstract formalizations of the types of claims that we analyze,  
and finally in Section~\ref{sec:execution} we compose the formalization of individual claims in such a way that a satisfying
assignment of the composed formula corresponds to a possible SPA execution.

\subsubsection{Formalization of Legal Facts}
\label{sec:facts}
An SPA according to the class diagram in Figure~\ref{fig:contract}
contains a set $\P$ of persons and a distinct set $\Os$ of objects.
In the formal model that we build 
we assume that the persons $p$ and the objects $o$ have unique identifiers.

Dates in an SPA are usually given by calendar dates.
Calendar dates are complicated to process due to the many
rules that govern them, such as the treatment of leap years.
In the formalization, we simplify the processing and
represent dates using integer variables.
An execution of an SPA begins on date $d_S=0$ with the signature of all contract
parties, referred to as the \textit{signing}. The \textit{closing} 
is performed on date $d_C$ when the transfer
of the business ownership is performed.
For each claim $c$ in an SPA, we define a date $d_c$
which represents its date of performance.

An ownership can only be transferred from the current owner to a new owner
if the ownership rights lie with the seller.
As depicted in Figure~\ref{fig:contract}, Property relations relating an object to an owner can be represented  
by associating the class \texttt{PropertyRight} using an association \texttt{Owner}
with a \texttt{Person} $p$, and using an association \texttt{Property} with an \texttt{Object} $o$. 
We define a set $\textit{\cal PR}$, consisting of a tuple $(p,o)$ for every
instance of \texttt{PropertyRight}, 
in the object diagram representing the contract instance that is being analyzed, c.f.\ Figure~\ref{fig:example}.
In our analysis model, we formally capture the property rights using a
first-order 
uninterpreted function predicate 
$\owner(\textit{Object}):\textit{Person}$:
\begin{align} 
\phi_{\owner} = \bigwedge_{(p,o)\in \textit{\cal PR}} \owner(o)=p.
\end{align} 
The first-order formula $\phi_{\owner}$ then represents the property relations
stated in the SPA.

\subsubsection{Formalization of Claims}
\label{sec:form_claim}
Having formalized the legal facts stated in an SPA
we are now ready to formalize the claims that it contains. 
Notice that we refer to classes, object instances, operations,  
association names and association roles as they are defined in 
Figure \ref{fig:contract} and Figure \ref{fig:example}.

We define a set $\Cs$ to consist of the claims stated in an SPA. 
The set $\Cs_I$ contains the \textit{primary claims} and
the \textit{independent warranty claims} of the considered SPA.
For every \textit{claim} $c$, the set $\Cs_c$ contains every
\textit{secondary claim} $s$ that is connected via the association 
\texttt{Trigger} to $c$.
In the following, we refer to a claim in $\Cs_c$
as a \textit{consequence claim} of $c$.

Each claim $c$ in the SPA contains a \texttt{Performance} $l_c$,
which specifies an action that must be executed for the claim to be performed.
The action to which $l_c$ refers is either represented by a logical formula or by the name of an
operation in Figure~\ref{fig:contract}.
In the latter case, $l_c$ is replaced by the logical formalization
of the pre- and postconditions of the corresponding operation.
For example, the precondition 
for a property transfer $p$ of an
object $o$ from the debtor $d$ to a creditor 
in terms of the property rights 
is represented by the constraint
$l_c^p\equiv\owner(o)=d$.

For every primary or secondary claim $c$ there exists a performance date $d_c$.
The value of $d_c$ is either $-1$, which we use to express 
the fact that the claim has not yet been performed,
or in the interval 
$[\duedate, \limitation]$. 
If \duedate\ is undefined in the considered SPA and, consequently, in the corresponding object diagram, then its default value is the value
of date \arise\ of the claim.
In case \duedate\ is a positive number, then
we replace it with $\arise+\duedate$.
In conclusion, we formalize a claim $c$ using the logical constraint  
\begin{align}
\phi_c \equiv (d_c=-1) \vee ((c.\duedate\leq d_c\leq c.\limitation) \wedge l_c).
\end{align}
In order to relate this to the pre-/postcondition semantics 
that we allude to above, notice that the subformula $c.\duedate\leq d_c\leq c.\limitation$ 
represents a precondition for the executability of $c$, and the subformula $l_c$ expresses the postcondition 
that $c$ has now been executed.

A \texttt{WarrantyClaim} $w$ is a specialization of a  \textit{secondary claim}
in the sense that it is an independent claim. 
For this reason, in our formalization, 
the constraint $d_w=-1$ encodes that $w$ is met, and otherwise 
it is breached on a date $d_w\geq 0$.
A warranty $w$ is formalized by 
\begin{align}
\phi_w \equiv (d_w=-1 \Rightarrow l_w)\vee (w.\duedate \leq d_w\leq w.\limitation).
\end{align}

If a \textit{primary claim} or an \textit{independent claim} is not performed,
the \texttt{Debtor} is obliged to perform a consequence claim $s$
that is associated with the \textit{claim}. 
A \textit{primary claim} is breached when it is not performed ($d_c=-1$),
while a warranty is breached on the date $d_w$ of the notification
that the \texttt{WarrantyCondition} is not met.
In order to formally capture the consequence claims that belong to a claim, we introduce
a fresh integer variable $d_c'$ with value $-1$ when the associated
\textit{claim} is performed.
The value of $d_c'$ is $c.\duedate$ for a breached \textit{primary claim}
and the value of $d_c$ for a breached warranty.
In case $s$ is a performance claim, its formalization with a performance
date $d_s$ and a \texttt{Performance} $l_c$ is defined as 
$\phi_s\equiv(d_c'< d_s\leq s.\limitation) \wedge l_c$.
In case $s$ is a restitution claim, the SPA is withdrawn,
formally $\phi_s\equiv d_c'< d_s\leq s.\limitation$.
In case $s$ is a compensation claim, the debtor pays a positive
compensation $l_s$ to the creditor.
The value of $l_s$ is the \texttt{Compensation} 
that specifies the amount of the compensation payment. 
As expressed by Formula~\ref{eq:damage}, $l_s$ 
is constrained to be in a range 
between a minimum amount \Min\ and a maximum amount \Max.
\begin{align} 
\phi_{\damage}\equiv l_s = \left \{ \begin{aligned}
    &0, && \textit{if } \damage <\Min \\
    & \Max && \textit{if } \damage>\Max \\
    & \damage, && \textit{otherwise}
  \end{aligned}\quad  \right .
  \label{eq:damage}
\end{align}  
In every of the above cases, the compensation is performed. 
In conclusion, the compensation claim is formally a constraint
\begin{align}
\begin{aligned}
\phi_s\equiv \phi_\damage\wedge 
((d_s=-1\Rightarrow \damage=0)\; \vee \\
(d'_c< d_s\leq s.\limitation))
.    
\end{aligned}
\end{align}

\subsubsection{Execution of an SPA.}
\label{sec:execution}

We are now prepared to present the logical formalization of the executions permitted by an SPA by a formula 
$\phi_\SPA$, given in Equation~\ref{eq:SPA}.
It captures the property rights and the claims included as an SPA 
by \textit{conjoining} their above given formalizations. It also ensures 
that in an execution of an SPA, for every \textit{primary}
and every \textit{independent claim}, either the claim
or one of its consequence claims will be performed.
\begin{align}
\phi_\SPA\equiv\phi_{\owner}\wedge
\displaystyle\mathop{\bigwedge}_{c\in \Cs}\;\phi_c \wedge
\displaystyle\mathop{\bigwedge}_{c\in \Cs_I}\;
(d_c\geq 0 \vee \bigvee\limits_{\forall s\in \Cs_c} d_s\geq 0).
\label{eq:SPA}
\end{align}
An SMT solver,
such as Z3,
will produce a satisfiable model for $\phi_{\SPA}$ 
in case such a satisfiable model and, consequently, an execution of the SPA exist.
This model then represents an execution of the SPA that is consistent with
all constraints specified in the SPA and formally represented by $\phi_\SPA$.

During the execution of a contract, the contracting parties often have a 
choice as to whether to fulfill a \textit{primary claim} or an \textit{independent claim}, 
or to breach that \textit{primary claim} or \textit{independent claim} 
and be subject to a \textit{secondary claim} that results from the 
breach. This second option may put a contracting party into an advantageous 
position (``efficient breach''). It is not the objective of our work to determine
advantageous courses of action for the contracting parties. Nonetheless, in the
analysis it might be worthwhile to determine whether an execution of the 
contract involving just \textit{primary claims} is at all possible, 
in order to obtain a 
deeper understanding of the functioning of the considered contract. 
To support this idea,
we propose a logical encoding that permits the prioritization of \textit{primary claims}
over \textit{secondary claims} during analysis.
We introduce a set
$\phi_{\textit{soft}}$ of logical assertions that is encoded in Z3 using 
what is referred to as 
``soft-asserts'', c.f.\ Equation~\ref{eq:soft}.
\begin{align} 
\phi_{\textit{soft}}\equiv\displaystyle\mathop{\bigwedge}_{c\in \Cs_I}\; d_c\geq 0
\; \wedge \displaystyle\mathop{\bigwedge}_{s\in \Cs_c}\; d_s=-1
\label{eq:soft}
\end{align} 
The SMT solver Z3 computes an optimal solution for the partially
satisfiable MaxSMT problem $\phi_{SPA}\wedge \phi_\textit{soft}$.
The computed solution is optimal in the sense that the smallest number 
of assertions in $\phi_{\textit{soft}}$, each corresponding to a breach
of a primary or independent claim and the execution of a secondary claim, 
is made satisfied, which means that the corresponding secondary claim 
will be executed. 
As a consequence, Z3 returns a satisfying model representing an execution of the SPA that
satisfies as few consequence and independent claims as possible.

\subsection{Formalization of the Bakery SPA}
We now illustrate the application of the formalization described above
to the Pretzel Bakery SPA case study.

\subsubsection{Legal Facts in the Bakery SPA}
The bakery SPA $B$ describes a set of persons 
$\P^B=\{\texttt{Eva}, \texttt{Chris}, \texttt{Bank}\}$ and
an object set $\Os^B=\{\texttt{Bakery},\texttt{Price}\}$
which includes the shares of the \texttt{Bakery} and the purchase price
\texttt{Price}.
The SPA also states that the ownership of 
\texttt{Bakery} is transferred by way of security to \texttt{Bank}, encoded 
as 
\begin{align}
\phi_{\owner}^B\equiv\owner(\textit{Bakery})= \textit{Bank}.
\end{align}

\subsubsection{Claims in the Bakery SPA}
Eva is obliged by the \texttt{TransferClaim} to transfer the
shares of the bakery on a day $d_u$.
She has to perform the \texttt{TransferClaim} on the day of
\texttt{Closing} ($28$). In case she misses that date,
\texttt{Chris} has a claim on her delivery of the shares. 
Eva can only perform the transfer if she
is the owner of \texttt{Bakery}.
These facts are captured using the constraint
\begin{align}
\hspace{-1mm}
\phi_{\textit{TransferClaim}}\equiv &\; (-1= d_u)\vee
((28\leq d_u)\wedge
(\owner(\textit{Bakery})=\textit{Eva})).
\label{eq:ueber}
\end{align}
Furthermore, Chris is obliged by the \texttt{PayClaim}
to pay the amount of \texttt{Price} to Eva.
The condition for the transfer is formally captured by the constraint 
$\owner(\textit{Price})= \textit{Chris}$.
The transfer is performed on a day $d_z$ and is due
on day $28$.
The formalization of the \texttt{PayClaim} is then given as 
\begin{align}
\hspace{-1mm}
\phi_{\textit{PayClaim}}\equiv&
(-1= d_z)\vee((28\leq d_z)\wedge
\owner(\textit{PurchasePrice})=\textit{Chris})).
\hspace{-1mm}
\label{eq:payment}
\end{align}
If one of these two \textit{primary claims} is not performed, then
the related \texttt{RestitutionClaim} (shorthand: Res.)
allows the respective creditor
on a date $\geq 0$ to withdraw from the SPA, formally captured by the formulas 
\begin{align}
\hspace{-0.1cm}
\phi_{\textit{Res.Purchaser}}\equiv\; & d_{\textit{Res.Purchaser}}=-1
\vee \textit{PayClaim.DueDate} < d_{\textit{Res.Purchaser}} \\
\text{and} \; \phi_{\textit{Res.Seller}}\equiv\; & d_{\textit{Res.Seller}}=-1
\vee \textit{TransferClaim.DueDate} < d_{\textit{Res.Purchaser}}.
\label{eq:back}
\end{align}

If the \textit{warranty claim} \texttt{PretzelWarranty} is breached,
then Chris notifies Eva of this breach on a day $d_g\geq 0$. Notice that 
$d_g=-1$ is used to encode that there is no indication of a non-performance and, 
therefore, the warranty \texttt{Condition}, which stipulates that $\textit{Pretzels}=10,000$, is met.
The formalization for the \texttt{PretzelWarranty} is then given as 
\begin{align}
\phi_{\textit{PretzelWarranty}}\equiv\;&
(d_g=-1\wedge Pretzels=10,000)\vee
(28\leq d_g\leq 28+14).
\label{eq:warranty}
\end{align}
In the bakery SPA, the warranty has
the consequence \texttt{Claim1} of type \texttt{PerformanceClaim}, 
which implies that the seller must perform the pretzel guarantee on a date $d_n$
in case the warranty is not met. This is encoded using the following formula:
\begin{align}
\phi_{\textit{Claim1}}\equiv\;&
(d_n=-1) \vee (d_g< d_n\leq d_g+28 \wedge
Pretzels=10,000).\label{eq:folge1}
\end{align}
On the other hand, following the idea of an efficient breach 
it may be more advantageous for the debtor
to pay a compensation $l_s$ on a date $d_s$ 
according to the attribute \texttt{Compensation} of the object \texttt{Claim2}
of type \texttt{CompensationClaim}.
The value of $l_s$ is constrained by the formula $\phi_\damage^s$.
If no compensation occurs ($d_s=-1$), then $l_s$ is $0$.
The formalization of the logical constraint encoding \texttt{Claim2} is
\begin{align}
\phi_{\textit{Claim2}}\equiv&\; \phi_\damage^s \wedge
((d_s=-1 \wedge l_s= 0) \;\vee 
(d_g< d_s\leq d_g+ 28+14)).
\label{eq:sequence2}
\end{align}
In order to concretize $\phi_\damage^s$, consider that the amount of 
compensation $l_s$ is either $0$, or lies in the
range of values between a minimum value \texttt{Min} and 
a maximum value \texttt{Max}.
In order for a compensation to be paid, $l_s$ must exceed the value of \euro $1,000$, while 
an upper limit \texttt{Max} is not specified. 
$l_s$ is calculated according to 
the constraint $\phi_\damage^s$ defined as follows: 
\begin{equation} 
\phi_\damage^s \equiv l_s = \\ \left 
\{ 
   \begin{aligned}
    & 0,  \textit{if} \; (10,000-\textit{Pretzels}/100)*1,000 \leq 1,000, \textit{and}\\
    & (10,000-\textit{Pretzels}/100)*1,000,
     \textit{otherwise.}
 \end{aligned}\quad   
 \right .
  \label{eq:damageB}
\end{equation}  

\subsubsection{Execution of the Bakery SPA}

The overall encoding of the claims in the bakery SPA formalized in Equation~\ref{eq:SPA} is given as
\begin{align}
\begin{aligned}
\phi_\SPA^B\equiv &\phi_{\owner}^B\wedge (\phi_{\textit{TransferClaim}}\vee
\phi_{\textit{Res.Purchaser}})\wedge(\phi_{\textit{PayClaim}}\;\vee\\
& \phi_{\textit{Res.Seller}})\wedge
(\phi_{\textit{PretzelWarranty}}\vee \phi_{\textit{Claim1}}\vee \phi_{\textit{Claim2}}).
\end{aligned}
\label{phibspa}
\end{align}
The formula~\ref{phibspa} encodes the freedom to perform a claim or one of its consequences 
by a \textit{disjunction} of the individual claim constraints.
In order to express that all claims or their consequence claims need to be performed, we \textit{conjoin} 
the subformulae related to the individual claims and the legal facts to form the constraint 
$\phi_\SPA^B$ that represents the encoding of all claims in the SPA.
Notice that an execution of the bakery SPA does not necessarily need to perform
every \textit{primary claim}, but can also perform 
the associated \textit{consequence claim} for some or all of the \textit{primary claims}.
However, a satisfying model of $\phi_{\SPA}^B$ should preferably represent the performance of a sequence of 
\textit{primary} or \textit{independent claims}.
This is formalized by the following formulas 
which are characterized as "soft-asserts" in the smtlib encoding handed over to the 
SMT solver:
\begin{align}
\begin{aligned}
\phi_{\textit{soft}}^B\equiv &\;
d_u\geq 0\wedge d_z\geq 0\wedge d_{\textit{Res.Purchaser}}=-1
\; \wedge \\  &
d_{\textit{Res.Seller}}=-1\wedge d_g=-1 \;\wedge\; d_n=-1.
\end{aligned}
\end{align}
The bakery SPA is formalized by the constraint 
$\phi_\SPA^B \wedge \phi_{\textit{soft}}^B$. Each satisfying model
of this constraint represents a possible execution of the bakery SPA.

\paragraph{Number of Contract Executions}
The steps of a contract execution can be distinguished into the performance of claims 
and into the assignment of variable values. Examples of the latter are 
the number of possible choices of prices and the days on which the claims are performed.
For instance, each choice of a price leads to a distinct execution. Since there are infinitely many possible prices to choose from, there are infinitely many executions.
Notice that in general due to the use of variables of type integer or real in the formalization, the number 
of resulting distinct contract executions is (countably) infinite.

\begin{figure}[htb]
\centerline{
\includegraphics[width=0.6\textwidth]{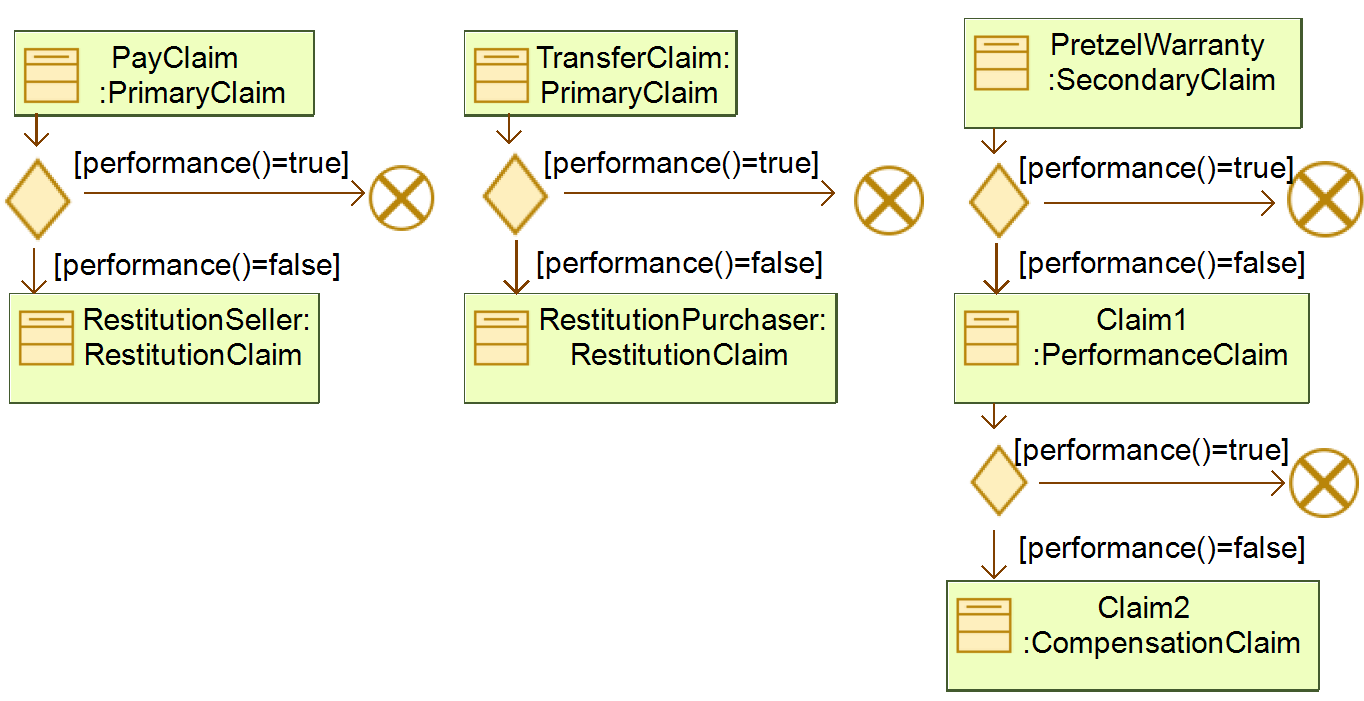}
}
\caption{Claims that are related by a \textit{Trigger} in the Bakery Contract.}
\label{fig:relations}
\end{figure}

We now argue why, nonetheless, only a finite number of different sequences
of claim executions need to be considered.
In a contract execution, every \textit{claim} is either performed or not. 
The claims are related by a \textit{Trigger}.
If a \textit{claim} is not performed, another \textit{claim} is triggered and arises.
The \textit{claims} of the bakery SPA trigger one another as depicted in the activity diagram in Figure~\ref{fig:relations}.
The claims that are related by a trigger \textit{Trigger} are in the same trigger set.
For the Bakery example, Figure~\ref{fig:relations} shows $3$ trigger sets with $2$, $2$ and $3$ elements.
Out of each trigger set, exactly one claim needs to be performed in an execution. 
This leads to $12$ different combinations of claim executions that can be performed 
in a contract execution.

\section{Contract Analyses}
\label{sec:verify}
In this section we discuss the syntactic and semantic consistency analyses that we 
propose, based on the above presented formalization of an SPA.

\subsection{Syntactic Analyses}
Our paper focuses on semantic analyses, based on our logical formalization.
However, in addition to these semantic analyses, our analysis framework
allows for the analysis of syntactic properties of an SPA. In particular, we 
can easily perform syntactic completeness checking, since the ontological
modeling allows us to infer some syntactic correctness and completeness
criteria, e.g. that an SPA lacks essential legal elements.
For example, an SPA is not legally valid if one of the essential elements
(\textit{essentialia negotii}) is missing.
According to the German civil code~\cite[§ 433]{BGB}, the 
contract of sale must contain the parties to the contract, the subject matter of the purchase and the purchase price.
In an SPA, every \textit{primary claim} also needs at least one consequential claim and every claim needs a \texttt{DueDate} by which it is to be performed.
Syntactic completeness checks can easily be performed when parsing
the block model of a concrete SPA instance, based on the Class diagram in Figure~\ref{fig:contract}.
However, in the sequel we will focus on presenting the semantic consistency
analysis of SPAs.

\subsection{Semantic Analyses}
Finding a dynamic inconsistency that occurs during SPA contract execution
requires reasoning about the dynamic semantics of executing an SPA.
As discussed above, an SPA describes both legal facts and claims agreed by the parties. 
For example, Eva agrees to transfer the shares of the bakery in the considered pretzel bakery SPA. 
In order to transfer ownership of the shares, she must be the owner of the bakery, and hence, the shares. 
This contradicts the fact that the bank owns the shares in the bakery. 
Interestingly, in spite of this semantic inconsistency 
the SPA can be executed or fulfilled because Chris can withdraw from the contract.
Nonetheless, an SPA which leaves Chris no option but to withdraw is neither meaningful
nor desirable and should therefore be tagged with a ``red flag'' as a result of the analysis.
This illustrates that for an SPA it is essential 
that a) each claim can be fulfilled
individually, and b) that at least one meaningful execution of the SPA exists.
Consequently, we propose two types of semantic consistency analyses for an SPA:

\newcommand{\one}[0]{I}
\newcommand{\all}[0]{II}
\begin{description}
    \item[\ Analysis \one:] Can each claim be performed?
      \label{it:one}
    \item[\ Analysis \all:] Does there exist an execution of the SPA?
      \label{it:all}
\end{description}

\subsection{Formalization of the Dynamic Consistency Analyses}

We now encode analyses \one\ and \all\ for a given SPA
using the formalized constraints presented in Section~\ref{sec:formal}.
If an encoded analysis is satisfiable, then an SMT solver, such as Z3, returns
a model given by a satisfying variable assignment.
The returned model represents either individual claims or entire 
contract executions that are consistent and can be executed.
If it is unsatisfiable, the SMT solver can return an
unsatisfiability core, which is a minimal subset of conditions
that contradicts satisfiability.
This subset indicates which claims in the SPA contradict each other
and, hence, cause an inconsistency in the SPA. The respective claims need to be changed
in the SPA in order to obtain a consistent contract.

Remember, a \textit{primary claim} $c$ is performed on date $d_c$
and a \textit{warranty} $w$ is asserted on day $d_w$.
In the following, in order to simplify the presentation we abuse notation and ignore that warranties and claims
behave differently.
In the formulae that follow, $d_c>=0$ needs to be substituted by $d_w==-1$ for every warranty $w$.

\paragraph{Analysis~\one~(Claim Consistency)} is encoded using the constraints
$\Phi_c$ and $\Phi_s$ for every claim in $\Cs$.
We presented the formalization for a claim $c$ by a constraint $\phi_c$, which
has to be performed on a date $d_c$ in Section~\ref{sec:form_claim}. 
For every claim, in order for it to be fulfillable, the property relation represented
by $\phi_{\owner}$ has to hold.
A \textit{primary claim} $c$ or an \textit{independent claim} $c\in \Cs_I$
can be performed when the following constraint is satisfiable:
\begin{align}
\Phi_c\equiv\phi_{\owner}\wedge \phi_c\wedge d_c\geq0.
\end{align}
A consequence claim $s\in \Cs(c)$ in an SPA usually needs to
be performed if, at the same time, the claim $c$ that is the \texttt{Trigger} of $s$
is not performed: 
\begin{align}
\Phi_s\equiv\phi_{\owner}\wedge (d_c=-1)\wedge\phi_c\wedge\phi_s\wedge d_s\geq0.
\end{align}
If every constraint $\Phi_c$ and $\Phi_s$ is independently satisfiable,
then every claim of the considered formalized SPA can be performed.

\paragraph{Analysis~\all~(Contract Executability)} checks whether an execution of 
the SPA exists by assessing the satisfiability of 
\begin{align}
\Phi_\SPA\equiv\phi_\SPA\wedge \phi_{\textit{soft}}.
\label{eq:SPA-soft}
\end{align}
In addition, Analysis~\all\ recognizes whether all 
\textit{primary claims} and \textit{independent claims} 
can be performed in the same execution, which means
that no \textit{secondary claim} in the contract needs to be executed.
If that is the case, then every constraint in $\phi_{\textit{soft}}$
is satisfied.

\subsection{Dynamic Consistency Analyses of the Bakery SPA}

We now apply the dynamic analyses to the pretzel bakery SPA.
The object diagram in Figure~\ref{fig:example} represents the 
information relevant for the analysis, such as the values of dates, 
the amounts of money and  the names of the seller and the purchaser.
After reading and parsing collection of blocks representing this 
contract, \textit{ContractCheck} generates the following analyses and
checks them for satisfiability by invoking the SMT solver Z3.

\paragraph{Analysis \one~(Claim Consistency)} checks whether the individual 
\textit{primary claims} and \textit{independent claims} are fulfillable.
The \textit{TransferClaim} is formalized in the Equation~\ref{eq:ueber}
and the corresponding Analysis \one\ is
\begin{align}
\begin{aligned}
\Phi_{\textit{TransferClaim}}\equiv
&\;\phi_\owner^B\;\wedge \phi_{\textit{TransferClaim}} \wedge\;
d_{u}\geq 0\\
\equiv&\;\owner(\textit{Bakery})= \textit{Bank}\;\wedge
(d_u=-1\;\vee\\
&(28\leq d_u\;
\wedge \owner(\textit{Bakery})=\textit{Eva}))\;\wedge\; d_u\geq 0.
\end{aligned}
\end{align}
$\owner$ is a function, therefore the two constraints $\owner(\textit{Bakery})=
\textit{Bank}$ and $\owner(\textit{Bakery})= \textit{Eva}$
are functionally inconsistent.
As a consequence, the claim cannot be \textit{performed}, which indicates
that the contract contains a logical inconsistency.
The SMT solver returns these two inconsistent constraints as an unsatisfiability core.

Analysis \one\ for the \textit{PayClaim} (Equation~\ref{eq:payment}) is performed by 
determining satisfiability of the following constraint:
\begin{align}
\begin{aligned}
\Phi_{\textit{PayClaim}}\equiv&\;\owner(\textit{Bakery})= \textit{Bank}\;
\wedge (d_z=-1\;\vee \;\\\
& (28\leq d_z \wedge \owner(\textit{Price})=\textit{Chris}))
\wedge d_z\geq 0.
\end{aligned}
\end{align}
The SPA states that the closing should be performed on day $d_z=28$.
A possible model that an SMT solver computes which satisfies this constraint contains
the assignments
$d_z=28$ and $\owner(\textit{Price})= \textit{Chris}$ which indicates that 
Chris pays the purchase price on day $28$ after signing.
We can conclude that the \texttt{PayClaim} can be \textit{performed}.

Analysis \one\ for the \textit{PretzelWarranty} (Equation~\ref{eq:warranty})
is  performed by 
determining satisfiability of the following constraint:
\begin{align}
\begin{aligned}
\Phi_{\textit{PretzelWarranty}}\equiv&
\owner(\textit{Bakery})= \textit{Bank}\;\wedge\;
(28\leq d_g\leq 28+14\;\vee\\
&(d_g=-1\wedge pretzels=10,000)) \wedge d_g=-1.
\end{aligned}
\end{align}
$\Phi_{\textit{PretzelWarranty}}$ is satisfiable for
$d_g=-1$ and $\textit{pretzels}=10,000$.
This means that the pretzel bakery meets the warranty requirement and
the warranty claim is performed.

\paragraph{The Analysis \all~(Contract Executability)} for the pretzel bakery example 
is encoded in the constraint $\Phi^B_\SPA\equiv \phi_\SPA^B \wedge \phi_{\textit{soft}}^B$, following
Equation~\ref{eq:SPA-soft}. It can be used to check whether an execution of the bakery SPA exists.
The SMT solver that we use in our automated analysis searches for a satisfiable assignment and computes,
for instance, a model with the following date assignments:
\begin{align}
\begin{aligned}
d_u=-1, d_z=28, d_g=-1,d_n=-1,d_s=-1,\\ d_{\textit{ResitutionSeller}}=-1,
d_{\textit{ResitutionPurchaser}}=29.
\end{aligned}
\end{align}
This outcome implies that the bakery SPA can be performed, even though it is inconsistent
since the \textit{primary claim} \texttt{TransferClaim} cannot be performed.
It exemplifies that the inconsistency of an SPA does not imply that it cannot be performed.
Furthermore, as the SMT solver confirms, if the bakery
were not assigned by way of security, but instead
$\owner(\texttt{Bakery})= \textit{Eva}$ held, 
the pretzel bakery SPA would be consistent.

\subsection{Further Dynamic Analyses}

We now present some further analyses that address single claims. They differ from Analysis I in the 
sense that they are not a prerequisite for the Analysis II.

\paragraph{Unsatisfiability of Single Claims}
A claim may or may not be satisfied during the execution of a contract. The model of a contract must support both cases. If a claim is not fulfilled, further claims arise as a consequence for the non-fulfillment. 
We propose a further analysis which establishes that a contract is executable even though 
a claim $c$ may not be satisfied, expressed by the constraint $\Phi_{\textit{unsat}\_c}$:
\begin{align}
\Phi_{\textit{unsat}\_c}\equiv&\; \phi_\SPA \wedge
c.\duedate=-1 .
\label{eq:unsat_c}
\end{align}
Assume the pretzel bakery example SPA not to specify any restitution claim. 
When we apply the analysis in Equation~\ref{eq:unsat_c} to the \texttt{TransferClaim}, 
the constraint $\Phi^B_{\textit{unsat}\_\textit{TranferClaim}}$ will be created:
\begin{align}
\Phi^B_{\textit{unsat}\_\textit{TranferClaim}}\equiv&\; \phi_\SPA \wedge d_u=-1 .
\end{align}
Since there is no consequence for the \texttt{PerformanceClaim},
no contract execution exists in which this claim is unsatisfied. It follows that
$\Phi^B_{\textit{unsat}\_\textit{TranferClaim}}$ is unsatisfiable.
One possible reason for this analysis to fail is that a primary or
independent claim does not possess a consequence claim. 
Notice that, if for instance a warranty is unsatisfied, a consequence like a compensation
should be paid.
Without a consequence claim, the encoding enforces that the claim is always
satisfied as there is no consequence claim that can be satisfied instead.

\paragraph{Claim Defense Analysis}
A claim $c$ can be defensed because of the lack of performance of another claim $d$. This relation is represented by a $\textit{Precede}$ association.
When $c$ would become due before $d$ and $d$ precedes $c$, then $c$ cannot be performed in time.
The analysis of the constraint $\Phi_{\textit{Precede}\_c\_d}$ checks for such 
timing inconsistencies.  
It is satisfiable if the inconsistency exists in the considered SPA: 
\begin{align}
\Phi_{\textit{Precede}\_c\_d}\equiv&\; \phi_{\textit{owner}}
\wedge\phi_c \wedge \phi_d \wedge c.\duedate< d.\duedate .
\end{align}
In the pretzel bakery SPA example, the payment of the bakery precedes the transfer occurring first.
Consequently, the due date of the \texttt{PayClaim} should not be after the 
due date of the \texttt{TransferClaim}, which is tested by checking the satisfiability of the 
formula 
$\Phi_{\textit{Precede}\_\textit{Transfer}\_\textit{Pay}}$:
\begin{align}
\Phi_{\textit{Precede}\_\textit{Transfer}\_\textit{Pay}}\equiv&\; \phi_{\textit{owner}}\wedge\phi_{\textit{Pay}} \wedge \phi_{\textit{Transfer}} \wedge d_{\textit{Transfer}}< d_{\textit{Pay}} .
\end{align}
Applying the analysis to the pretzel bakery SPA example, the constraint
which indicates that there is no such inconsistency between the \texttt{TransferClaim} 
and the \texttt{PayClaim}. By changing the due date of the payment to $29$,
$\Phi_{\textit{Precede}\_\textit{Transfer}\_\textit{Pay}}$ becomes satisfiable with $d_\textit{Transfer}=28$ and $d_\textit{Pay}=29$, which 
means that an inconsistency has been detected.

\paragraph{Limitation Analysis}
The \textit{due date} and \textit{limitation} of a claim
can refer to other time events in the contract.
In the pretzel bakery SPA, for instance, the \texttt{DueDate} of \texttt{Claim1} is $+28$.
The relative time reference means that \texttt{Claim1} has to be fulfilled
within $28$ days after the assert date $d_g$ of \texttt{PretzelWarranty}.

\begin{figure}[htb]
\centerline{
\includegraphics[width=0.5\textwidth]{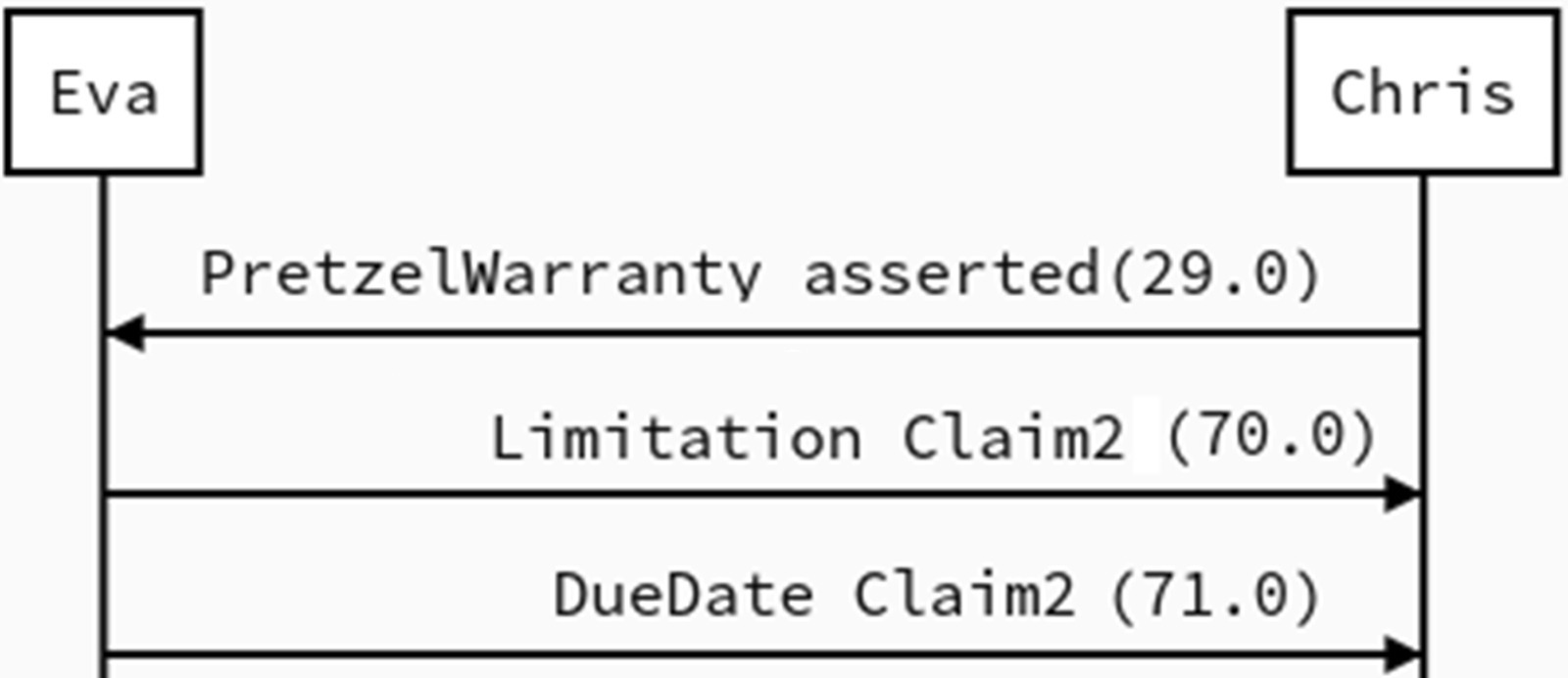}
}
\caption{Sequence diagram computed by \textit{ContractCheck} illustrating a limitation error.}
\label{fig:limitation}
\end{figure}

The combination of several \textit{due dates}
and \textit{limitations} can result in an inconsistency
where, because of its \textit{limitation} date, 
the \textit{due date} of a claim is later than the date at which 
the claim expires.
Analysis $\Phi_{\textit{Limitation\_c}}$ checks this inconsistency
for every \textit{claim} $c\in\Cs$ with a \texttt{Limitation}.
For this analysis, we need to (textually) substitute $c.\duedate\leq d_c\leq c.\limitation$
by $c.\duedate\leq d_c$, which we indicate by the notation $c.\duedate\leq d_c/c.\duedate$ and the
enclosure of the formula to which the substitution applies in squared backets:
\begin{align}
\begin{aligned}
\Phi_{\textit{Limitation}\_c}\equiv&\; \phi_\SPA[c.\duedate\leq d_c/c.\duedate\leq d_c\leq c.\limitation]\\ &
\wedge\;c.\textit{Limitation} < c.\textit{DueDate} .
\end{aligned}
\end{align}
In the bakery SPA, only the claims \texttt{Claim1}
and \texttt{Claim2} contain a \texttt{Trigger}
to \texttt{Pretzel\-Warranty} and a
\texttt{Limitation} of $70$ days.
Hence, the limitation analysis is only created for these two claims.  
The SMT solver Z3, invoked by \textit{ContractCheck}, computes that $\Phi_{\textit{Limitation\_Claim1}}$
is unsatisfiable, which entails that  the time constraints are consistent.
For $\Phi_{\textit{Limitation\_Claim2}}$, Z3 computes a model
that, for instance, contains the following assignments:
$d_{\textit{Claim1}}=-1$, $d_{\textit{Pretzel\-Warranty}}=30$ and
$d_{\textit{Claim2}}=71$.
With this contract execution, illustrated by the Sequence Diagram in Figure~\ref{fig:limitation},
Chris asserts the \texttt{Pretzel\-Warranty}
on day $29$, then Eva tries to perform \texttt{Claim1} 
for $28$ days but fails.
Now, Eva has to pay a compensation that is due after another $14$ days.
In summary, the compensation claim is due after $71$ days,
even though, due to the \texttt{Limitation}, 
the legal basis of the claim is outdated after $70$ days.
This shows that there is an inconsistency in the
timing between the due dates and the \texttt{Limitation} of the
\texttt{PretzelWarranty}.

\section{The \textit{ContractCheck} Tool}\label{sec:tool}

\subsection{\textit{ContractCheck} Workflow}

An SPA is usually formulated in natural language and not formalized.
Natural Language Processing to automatically obtain a semantic model 
of the SPA is beyond the scope of this paper.
Instead, we provide a collection of parameterized structured English text blocks to the user that 
can be selected and properly parameterized in order
to reflect the meaning of the SPA to be analyzed.
The \textit{ContractCheck}~\citep{CCTool} tool
for the consistency analysis of SPAs that 
we describe in this section parses these text blocks
and translates them into an internal formal
representation. In doing so, it gives these text
blocks a formal semantics. This relieves the user
from the task to provide a semantics, and thereby
a formalization, in a manual fashion.

\begin{figure}[htb]
\centerline{
\includegraphics[width=0.7\textwidth]{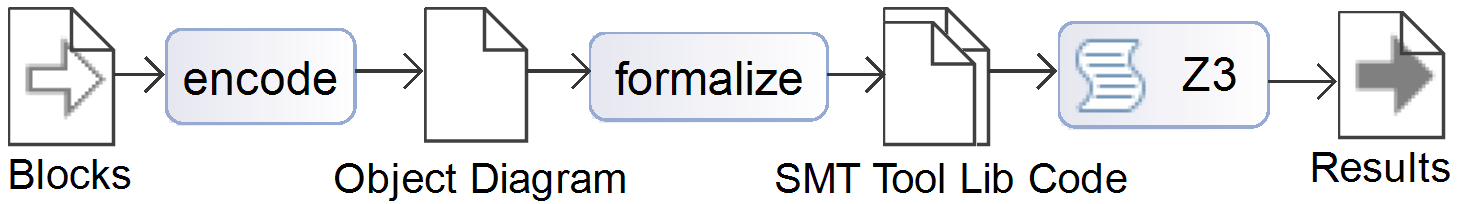}
}
\caption{\textit{ContractCheck} analysis workflow, modeled using the BPMN notation~\citep{BPMN}.}
\label{fig:tool}
\end{figure}

The workflow of the tool \textit{ContractCheck} is depicted in the diagram in Figure~\ref{fig:tool}.
After the user has selected and parameterized the
text blocks, \textit{ContractCheck} will compile
a UML object diagram from the text blocks and the
information regarding the class structure encoded
in the SPA class diagram (see 
Figure~\ref{fig:contract}). Note that the object 
diagram merely serves as an internal representation 
and does not need to be edited by the user. 
\textit{ContractCheck} extracts the formal 
representation of the SPA to be analyzed from the 
object diagram and synthesizes the logical 
constraints that we presented in 
Section~\ref{sec:formal} in \textit{smtlib2}-format.
Next, \textit{ContractCheck} invokes the Z3 SMT 
solver to perform the consistency analyses described in Section~\ref{sec:verify}. Finally, 
\textit{ContractCheck} translates the analysis 
results computed by Z3 into user-readable format and produces the results in its user interface.

\subsection{Text Blocks in \textit{ContractCheck}}

\begin{figure}[htb]
{\footnotesize 
\begin{tabular}{ l l}
ID:& Block1\\
\hline
Text:& \makecell[l]{The seller \$seller.Name hereby sells shares of \$shares.Name,
with all\\
rights and obligations pertaining thereto, to the Purchaser\\
\$purchaser.Name, who accepts such sale.}\\
\hline
Object:& \makecell[l]{``spa:SPA'', ``seller:Person'',``purchaser:Person'', ``share:Share'',\\ ``transfer:PrimaryClaim''}\\
\hline
Assignment:& \makecell[l]{``seller.name=Eva'', ``purchaser.name=Chris'',
``spa.Seller=\$seller'', \\
``transfer.Performance=\$shares.transfer(\$purchaser)'',
\ldots
}
\end{tabular}
}
\caption{Excerpt from Text Block Encoding of Bakery SPA in JSON format}
\label{fig:textblock}
\end{figure}

In \textit{ContractCheck}, text blocks serve to capture
the semantics of the different clauses in an SPA
in a structured fashion. Figure~\ref{fig:textblock}
gives an excerpt of the text block that represents
a portion of the clause 1.1 of the pretzel bakery SPA that can be found in full length in the Appendix.
A text block
has a unique identifier (\textit{ID}) that allows one to
refer to it. The \textit{Text} field contains a 
parameterized structured natural language 
description of the legal content of the 
contract clause that the text block represents.
The \textit{Object} field provides references to 
the objects that the text block is parameterized with.
The \textit{Assignment} field gives concrete values
for the object parameters that are referenced in 
the text block.
The information contained in the \textit{Object}
and \textit{Assignment} fields is then used by 
\textit{ContractCheck} to compile the object 
diagram used as an internal semi-formal representation
of the SPA, c.f.\ Figure~\ref{fig:example}.

More concretely, 
the text block with the ID \textit{Block1} 
given in Figure~\ref{fig:textblock} 
defines the essential
components of an SPA: A seller, a purchaser, the shares to be sold and a price.
The \$-character in the text indicates the assignment of a variable value,
such as the attribute \texttt{Name} for a person.
For instance, the assignment to \textit{\$seller.name} is
currently \textit{Eva}, as defined in the \textit{Assignment} section of the block.
A block may reference the variables of another block by using the \$-character. 
For instance, \$Block1\_share refers to the variable \textit{share} in \textit{Block1}.
The instance \textit{Bakery} of type \texttt{Shares} is defined in another block.
The inter-block reference semantics is necessary to
assign \$Block1\_share to the attribute \texttt{Property} of \textit{Bakery}.

We envision two usage scenarios for \textit{ContractCheck}, see also Figure~\ref{fig:approach}. One scenario is based on the
idea that a comprehensive library of text blocks
will be developed, so that existing textual SPAs 
can be manually mapped to blocks chosen from 
this library. 
The parameters of the text blocks will then
be filled with concrete values from the textual
contract at the time when this mapping is performed. This modeling step is supported
by the user interface of \textit{ContractCheck}.
The second scenario is based on the idea that the
contract will be defined using a set of text 
blocks from the library that represents the clauses
and stipulations of the SPA. A contract text can 
then be synthesized by \textit{ContractCheck} as
a by-product of the analysis. This second scenario
fits well into the increasingly popular use of 
contract synthesis or generation tools in legal practice. The 
consistency analysis performed by 
\textit{ContractCheck} is agnostic with respect to
the use of either of these two usage scenarios.

\subsection{\textit{ContractCheck} Analysis Result Presentation}

The analysis results are displayed in the 
user interface of \textit{ContractCheck}.
The results of the Analyses I and II and the further dynamic analyses 
are either satisfying models, which provide value assignments to the logical
variables which render the analysis problem satisfiable, or unsatisfiability cores
which give value assignments that render the analysis problem unsatisfiable.

\begin{figure}[htb]
\centerline{
\includegraphics[width=0.5\textwidth]{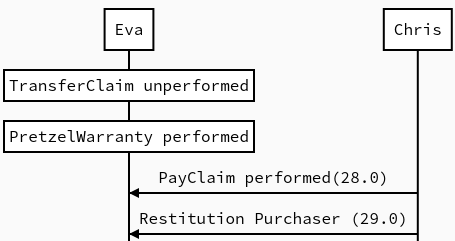}
}
\caption{Bakery SPA execution visualized in the form of a UML Sequence Diagram}
\label{fig:exe}
\end{figure}

In case a satisfying model is returned, the valuations
of the variables in the model indicate possible 
dates in which claims in the contract are 
executed. 
\textit{ContractCheck} depicts such a 
satisfying model by a sequence diagram.
For the bakery SPA, a computed execution is depicted in Figure~\ref{fig:exe}. It indicates that
the \texttt{TransferClaim} is unperformed by Eva,
the \texttt{PretzelWarranty} is performed by her,
Chris performs the \texttt{PayClaim} on day 28 
after signing, and is compensated for this payment
by a \texttt{Restitution} on day 29. This execution
scenario illustrates that the contract can be 
executed in spite of the inconsistency regarding
the ownership of the bakery, which means that the
primary claims cannot be executed.

\begin{figure}[htb]
\centerline{
\includegraphics[width=\textwidth]{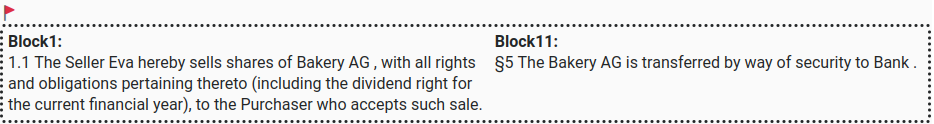}
}
\caption{Inconsistent output for the Bakery SPA.}
\label{fig:result1}
\end{figure}

In case the SMT analysis invoked by \textit{ContractCheck} returns a result of unsatisfiability of the constraint system, an
unsatisfiability core will be returned. 
It contains a maximum number of conjunctively 
connected logical constraints
so that any true subset of it is still satisfiable.
The constraints contained in the unsatisfiability core 
are associated with certain text blocks, which \textit{ContractCheck} ``red flags'' and 
draws side by side in the user interface, as shown in Figure~\ref{fig:result1}. 
In this case, the \textit{ContractCheck} analysis
result reveals the contradiction between the 
claimed pretzel bakery ownership by Eva in 
Block1, and the stated transfer of the pretzel bakery ownership to the bank in Block 11.
These are the text blocks that created 
a set of unsatisfiable constraints contained in the
unsatisfiability core of $\Phi_{\textit{TransferClaim}}$.
The depicted claims help the user to identify the claims that contradict each other and, hence, to 
``debug'' the considered SPA.

\begin{figure}[htb]
\centerline{
\includegraphics[width=8cm]{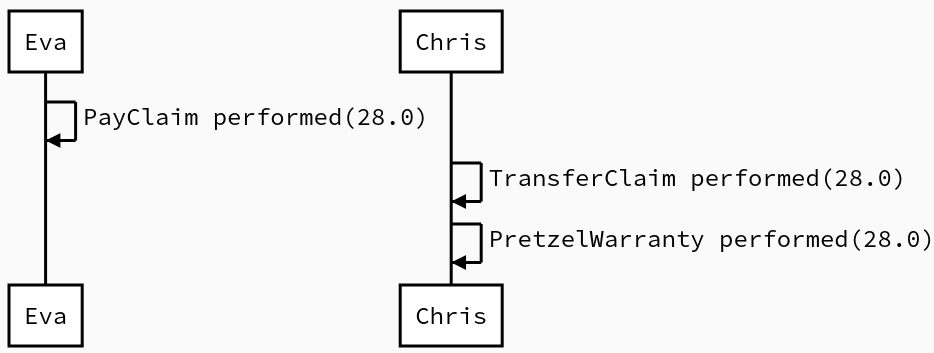}
}
\caption{Execution of the Bakery SPA without inconsistencies.}
\label{fig:result2}
\end{figure}

In order to test our approach, we created an instance of the Bakery example without inconsistencies.
Namely, the bank is no longer owner of the bakery and we also set the warranty to a limitation of 56 days.
The tool does not find any inconsistency and depicts the execution in Figure~\ref{fig:result2}.

\section{Case Study}\label{sec:evaluation}

To evaluate {\em ContractCheck} on an SPA of realistic size 
we have applied it to a slightly modified version of an SPA-template taken 
from~\cite{MeyerSparenberg.2022b}. This contract template is bilingual in German and English.
It consists of twelve paragraphs. The combined German and English text
spans 21 pages of a Word document and contains about 11.300 words. 

 We used Meyer-Sparenberg's original template as a starting point and adapted it by incorporating provisions commonly found in a wide range of standard contracts, which yielded the contract text that we use as a case study in this section. To identify these common clauses, we reviewed the sample contracts provided in 
the form books~\cite{Hoyenberg.2020,
Hoyenberg.2020b, MeyerSparenberg.2022a, MeyerSparenberg.2022b,
Pfisterer.2022, Seibt.2018a, Seibt.2018b, Seibt.2018c, Seibt.2018d}.
A quantitative analysis based on matching headings in the case study contract 
to headings in the form book sample contracts revealed that the clauses 
included in the case study contract also appear in between 75\% and 100\% of 
the contracts provided by the considered form books. It is safe to assume that
a large portion of the SPAs used in practice are based on these form book samples,
because, to our knowledge, these form books are the most commonly used in practice in Germany.
We consequently assume that our case study reflects a wide range of standard clauses and provisions, making it a representative test case for ContractCheck.

In order to be able to analyze the capabilities of 
{\em ContractCheck} to detect inconsistencies in 
contract texts it is necessary to seed inconsistencies
into these texts~\citep{5487526}. 
In particular, the following five inconsistencies were seeded:
\begin{description}

\item[a.] The contract states that the seller owns 100\% of the company shares, but a warranty clause was modified to state that the seller owns only 90\% of the shares. This creates an inconsistency between the actual share ownership specified and what is warranted.

\item [b.] The minimal compensation of a compensation claim is $20.000$\euro. The sum of all compensations  is restricted by a limitation of liabilities to $1\%$ of the purchase price of $1.800.000$. As a consequence, no compensation can ever be paid.

\item[c.] A clause was added stating that non-existing real estate properties used by the company have 
no third party rights. However, the contract specifies that the company owns no real estate property, 
so this clause is inconsistent.

\item[d.] The clause regulating payment of an additional purchase price upon adoption of the company's annual financial statements was modified to remove the requirement that the statements be transmitted to the seller. This creates a potential inconsistency where payment could become due before the seller receives the statements.

\item[e.] The effective date for accrual of interest on the purchase price was falsely specified as a date that comes after the stipulated latest possible closing date. This results in an impossibility where interest would accrue after the date the price is paid.

\end{description}
As we shall illustrate next, all of these seeded faults can be
detected by \textit{ContractCheck} using one of the defined semantic consistency analyses.

\begin{table}[htb]
{\centering
\begin{scriptsize}
\begin{tabular}{|c||c|c|r|r|r|}
\hline
\multicolumn{1}{|c||}{Analysis} &
\multicolumn{1}{c|}{\textit{\#Err/\#An}} & 
\multicolumn{1}{c|}{Time} & 
\multicolumn{1}{c|}{Memory} & 
\multicolumn{1}{c|}{\textit{\#Var}} & \multicolumn{1}{c|}{\textit{\#Con}} 
\\
\hline \hline
Contract Executability (\all) & 0/ \  1 & 0.030s & 3.17MB & 141 & 491\\
Claim Consistency (\one) & 5/47 & 0.005s & 1.45MB & 21 & 48 \\
Claim Unsatisfiable & 1/47 & 0.022s & 3.02MB & 138 & 312 \\
Claim Defense& 1/ \ 1 & 0.002s & 1.32MB & 19 & 30\\
Limitation Check & 1/13 & 0.017s & 3.16MB & 141 & 495\\
\hline
\end{tabular}
\end{scriptsize}
\vspace{0.2cm}
\caption{\label{tab:result_model} Quantitative experimental results for the analysis of a realistic size contract. Maximal result values for individual analyses in each of the analysis categories (rows.} }
\end{table}

\paragraph{Analysis Results}
We have applied the analyses implemented in {\em LegalCheck} to 
the inconsistency-seeded realistic size SPA contract described above. 
From a qualitative perspective, all seeded inconsistencies were detected
by {\em LegalCheck}.

An overview of 
the quantitative results can be seen in Table~\ref{tab:result_model}.
For the different types of analyses mentioned in column \textit{Analysis}, 
\textit{ContractCheck} created a number \textit{\#An} of SMT analyses, 
and found a number \textit{\#Err} of errors in the SPA.
A particular type of consistency analysis encoded as an SMT problem contains at most 
a number \textit{\#Var} of variables and \textit{\#Con} of constraints.
In other words, these two columns indicate the maximum size of a constraint
system solved by the SMT solver in any problem in a particular row of the table.
For each type of analysis, the column \textit{Time} indicates the maximum computation
time and the column \textit{Memory} the maximum amount of consumed main memory  
that it took to check satisfiability of any problem in that row of the table.

The analysis type \textit{Contract Executability (Analysis II)} returned one execution of the SPA which shows that the SPA is executable.
The analysis type \textit{Claim Consistency (Analysis I)} created $47$ SMT satisfiability analysis problems and 
succeeded in detecting the seeded error $a$ and $4$ compensation claims that can never be paid because of the seeded inconsistency $b$.
The analysis type \textit{Claim Unsatisfiable} created $47$ SMT satisfiability analysis problems and 
succeeded in detecting the seeded error $c$.
The analysis type \textit{Claim Defense} created $1$ SMT satisfiability analysis problems and 
succeeded in detecting the seeded error $d$.
The analysis type \textit{Limitation Check} created $13$ SMT  satisfiability analysis problems 
and 
succeeded in detecting the seeded error $e$.
For every analysis, the memory demand of the SMT-solver was  below $3.17$MB, and satisfiability was decided within at most $30$ms.
These results show that our encoding of the logical conditions allows for an efficient SMT-based satisfiability analysis 
for SPAs of realistic size.

 In conclusion, the case study shows that the proposed method is effective, i.e., that it finds inconsistencies in a more complex contract, and that is its efficient, i.e., that the formalization and the SMT solving scale to a contract of more realistic size and complexity.

\section{Threats to Validity}
\label{sec:threats}

We view the following points as main threats to the validity of our results:
\begin{itemize}
    \item It is possible that there are mismatches between the intended semantics of
    a contract text and the semantics entailed by our formalization. We believe that 
    such a mismatch is rather unlikely since the formalizations of the different types 
    of claims have been carefully 
    reviewed by the legal experts in the author team of this paper.
    \item There is a chance that the reasoning engine used in our paper is inconsistent.
    Again, the chances of this happening are rather low since the Z3 SMT solver that we
use is in heavy academic and industrial use, for instance within Microsoft Corp.\ \citep{DBLP:conf/fm/Bjorner18}, and can consequently be considered to be well tested
and stable.
    \item It may be that the set of blocks that we propose in this paper does not 
    correspond well to the types of stipulations encountered in practical use, i.e., that our blocks
    do not generalize well. 
    As we argue in Section~\ref{sec:evaluation}, the clauses used in our case study of complex contracts
    generalize well to a large set of sample contracts
    from form books widely used in practice in Germany. We maintain
    that this implies that a fair number of blocks that we developed for our case 
    studies can be reused in the formalization of a large number of SPAs that
    can be found in practice.
        \item At this point it is unknown which portion of the types of stipulations that
    occur in practical contracts are covered by the blocks for which we have proposed 
    a formalization and an analysis. We assume that over time a large library of reusable blocks 
    will be developed so that coverage will steadily increase. Notice that we do not 
    claim completeness in the sense that all inconsistencies in a given SPA contract
    text will be identified -- we merely stipulate that our analysis is complete with
    respect to the set of blocks used in the analysis of a given contract and the
    formalizations proposed for this set of blocks.
\end{itemize}

\section{Conclusion}\label{sec:conclusion}

We have presented a method for the logical modeling and consistency analysis of
legal contracts using an SPA as an example. We provided an ontology for SPAs using UML class diagrams
and illustrated the refinement of this  ontology to a UML object diagram as a 
case study. We discussed the logical formalization of the SPA using decidable 
fragments of FOL via SMT solving. Finally, we introduced the tool
\textit{ContractCheck}, which allows the textual editing of contracts using building 
blocks, performs the automated derivation of the logical encoding of the contract
and the consistency conditions, invokes the Z3 SMT solver, and returns the analysis 
results to the user. 
We have evaluated the consistency analyses that \textit{ContractCheck} can perform on an SPA of realistic
size and noted that significant seeded inconsistencies 
in the contract could be discovered.

Future research will increase the size and complexity of the contract artifacts that 
we consider. We will further develop the analysis of the dynamic execution of 
contracts by introducing state-machine models, among other things, in order to assess the benefits of the contract for 
different contracting parties in the light of the possible dynamic execution scenarios. 
We are also currently working on an approach relying on 
machine-learning based Natural Language Processing that 
automatically maps natural language SPA texts to the 
blocks that we describe in this paper.

We see this work as an innovative contribution to improving the quality of complex legal artifacts using logic-based, 
automated analysis methods. 

\bmhead{Acknowledgements} We wish to thank Raffael Senn and David Boetius for
helpful comments on an earlier version of this paper.

\begin{appendices}
\pagebreak
\section*{Appendix}\label{sec:lexicon}

\subsection*{Text of the Pretzel Bakery SPA}

\begin{quote}

\textbf{§1 Main Content}

1.1 The Seller Eva hereby sells the shares of Bakery AG with all rights and obligations pertaining thereto (including the dividend right for the current financial year), to the Purchaser Chris who accepts such sale.

1.2 The Purchaser pays the purchase price 40.000\euro\ to the Seller.

1.3 If the transfer is not performed, the Purchaser has the right to withdraw.

1.4 If the payment is not performed, the Seller has the right to withdraw.
\\
\textbf{§2} The Seller hereby represents and warrants to the Purchaser in the form of an independent guarantee pursuant to Section 311 (1) of the German Civil Code and exclusively in accordance with the provisions of this Agreement that the following statements (the ``Warranties“) are true and correct as of the date of this agreement and that the Warranties set forth in this paragraph will also be true and correct as of the Closing Date:

2.1 The company can produce at least the 10.000 of Pretzels every day (Pretzel Warranty).
In case of the breach of the Warranty, it needs to be asserted within 14 days.
\\
\textbf{§3} The Purchaser’s rights arising from any inaccuracy of any of the Warranties contained in §1 shall be limited to supplementary performance claims and compensation claims against the Seller, subject to the provisions of

3.1 In case the Pretzel Warranty is not met and then the creditor may demand subsequent performance within 28 business days from the debtor after having transfered the shares.

3.2 In case the Pretzel Warranty is not met and the damage is above 1.000\euro\ then a compensation of 100\euro\ per 100 pretzels not baked pretzels is paid within 14 business days.
\\
\textbf{§4} Claims of §3 expire after 42 business days.
\\
\textbf{§5} The Bakery AG is transferred by way of security to Bank B.
    
\end{quote}

\subsection*{Text Blocks for the Pretzel Bakery SPA}

{\footnotesize 
\begin{tabularx}{\textwidth}{l l}
ID:& Block1\\
\hline
Text:& \parbox{10.3cm}{The seller \$seller.Name hereby sells shares of \$shares.Name,
with all rights and obligations pertaining thereto (including the
dividend right for the current financial year), to the Purchaser
\$purchaser.Name, who accepts such sale.}\\
\hline
Object:& \parbox{10.3cm}{
"spa:SPA", "seller:Person", "purchaser:Person", "shares:Shares",
"transfer:PrimaryClaim"}\\
\hline
Assignment:&\parbox{10.3cm}{
"purchaser.Name=Chris", "seller.Name=Eva", "spa.Seller=\$seller",
"spa.Purchaser=\$purchaser",
"shares.Name=Bakery AG", 
"spa.Object=\$shares", "spa.Claim=\$transfer", "spa.Closing=28",
"transfer.Performance=Bakery.transfer(\$purchaser)",\\
"transfer.Debtor=\$seller", 
"transfer.Creditor=\$purchaser", 
"transfer.DueDate=28"
} 
\end{tabularx}
\ \\
\ \\
\begin{tabularx}{\textwidth}{l l}
ID:& Block2\\
\hline
Text:& \makecell[l]{The purchaser pays the purchase price \$price.Amount \euro\ to the seller\\
on date \$payment.DueDate.}\\
\hline
Object:& \makecell[l]{ "spa:\$SPA", "price:PurchasePrice",
"payment:PrimaryClaim"}\\
\hline
Assignment:&\parbox{10.3cm}{
"spa=\$Block1\_spa",
"spa.Price=\$price",
"price.Amount=40000",
"payment.Debtor=Block1\_Purchaser",
"spa.Claim=\$payment",
"payment.Creditor=Block1\_Seller",
"payment.DueDate=28",
"payment.Performance=price.transfer(\$seller)"
} 
\end{tabularx}
\ \\
\ \\
\begin{tabularx}{\textwidth}{l l}
ID:& Block3\\
\hline
Text:& \makecell[l]{If the \$claim is not performed, the \$withdraw.Creditor\\
has the right to withdraw.}\\
\hline
Object:& "claim:\$Claim",	"withdraw:RestitutionClaim" \\
\hline
Assignment:& \makecell[l]{
"claim=\$Block1\_transfer",	"withdraw.Name=Restitution Purchaser",	\\
"withdraw.Trigger=\$claim", "withdraw.Debtor=\$claim.Creditor",\\ "withdraw.Creditor=\$claim.Debtor"
}
\end{tabularx}
\ \\
\ \\
\begin{tabularx}{\textwidth}{ l l}
ID:& Block4\\
\hline
Text:& \makecell[l]{If the \$claim is not performed, the \$withdraw.Creditor\\
has the right to withdraw.}\\
\hline
Object:& "claim:\$Claim",	"withdraw:RestitutionClaim" \\
\hline
Assignment:& \parbox{10.3cm}{
"claim=\$Block2\_payment",	"withdraw.Name=Restitution Seller",
"withdraw.Trigger=\$claim", 
"withdraw.Debtor=\$claim.Creditor", "withdraw.Creditor=\$claim.Debtor"
}
\end{tabularx}
\ \\
\ \\
\begin{tabularx}{\textwidth}{ l l}
ID:\hspace{1cm}\ \ & Block5\\
\hline
Text:& \parbox{10.3cm}{The Seller hereby represents and warrants to the
Purchaser in the form of an independent guarantee pursuant to Section 311
(1) of the German Civil Code and exclusively in accordance with the
provisions of this Agreement that the following statements (the
``Warranties“) are true and correct as of the date of this Agreement 
and that the Warranties set forth in this paragraph will also be true
and correct as of the Closing Date:}\\
\end{tabularx}
\ \\
\ \\
\begin{tabularx}{\textwidth}{ l l}
ID:& Block6\\
\hline
Text:& \parbox{10.3cm}{
The company can produce at least the \$amount of \$thing every day.
In case of the breach of the warranty, it needs to be asserted within \$warranty.Limitation days.}\\
\hline
Object:& \parbox{10.3cm}{"warranty:WarrantyClaim", "count:Integer", "amount:Integer",
"thing:String"}\\
\hline
Assignment:& \makecell[l]{"warranty.Name=PretzelWarranty",
"warranty.Debtor=\$Block1\_seller",\\
"warranty.DueDate=\$Block1\_spa.Closing", "thing=Pretzels", \\
"warranty.Creditor=\$Block1\_purchaser",
"warranty.Limitation = +14",\\
"warranty.Performance=(Block6\_count=Block6\_amount)",
"amount=10000",\\
"Block1\_spa.Claim=\$warranty"
}\\
\end{tabularx}
\ \\
\ \\
\begin{tabularx}{\textwidth}{ l l}
ID:& Block7\\
\hline
Text:& \parbox{10.3cm}{
The Purchaser’s rights arising from any inaccuracy of any of the Warranties contained in \$block
shall be limited to supplementary performance claims and compensation claims against the Seller, subject to the provisions of}\\
\hline
Object:&\parbox{10.3cm}{ "claim:\$Claim", "per:PerformanceClaim", "block:\$Block"}  \\
\hline
Assignment:&\makecell[l]{
"block=\$Block6",
``claim=\$Block6\_warranty", ``per.Trigger=\$claim",\\
``per.Name=Claim1", ``per.DueDate=+28",\\
``per.Debtor=\$claim.Debtor", ``per.Creditor=\$claim.Creditor"
} \\
\end{tabularx}
\ \\
\ \\
\begin{tabularx}{\textwidth}{ l l}
ID:& Block8\\
\hline
Text:& \parbox{10.3cm}{
In case the \$claim is not met and then the creditor may demand
subsequent performance within \$per.DueDate business days
from the debtor after having transfered the shares.}\\
\hline
Object:&\parbox{10.3cm}{ "claim:\$Claim", "per:PerformanceClaim"}  \\
\hline
Assignment:&\makecell[l]{
``claim=\$Block6\_warranty", "per.Name=Claim1",	"per.Trigger=\$claim",\\
"per.DueDate=+28", "per.Performance=\$claim.Performance",\\
"per.Debtor=\$claim.Debtor", "per.Creditor=\$claim.Creditor"
} \\
\end{tabularx}
\ \\
\ \\
\begin{tabularx}{\textwidth}{ l l}
ID:& Block9\\
\hline
Text:& \parbox{10.3cm}{
In case the \$claim is not met and the damage is above \$comp.Min \euro
then a compensation \$claim.Performance is paid
within \$comp.DueDate days.}\\
\hline
Object:&\parbox{10.3cm}{ "claim:\$Claim", "comp:CompensationClaim"}  \\
\hline
Assignment:&\makecell[l]{
``claim=\$Block6\_warranty", "comp.Name=Claim2", "comp.Min=1000",\\
"comp.DueDate=+42","comp.Trigger=\$claim",\\
"comp.Compensation=((Block6\_amount-Block6\_count)/100)*1000",\\
"comp.Debtor=\$claim.Debtor", "comp.Creditor=\$claim.Creditor"
} \\
\end{tabularx}
\ \\
\ \\
\begin{tabularx}{\textwidth}{ l l}
ID:& Block10\\
\hline
Text:& \parbox{10.3cm}{
Claims in \$block expire after \$d business days.}\\
\hline
Objects:& "claim:\$Claim", "d:Date", "block:Block"\\
\hline
Assignment:& \makecell[l]{
``block=Block8", ``d=28+42", ``\$\{//\$block//Claim\}.Limitation=\$d"
}
\end{tabularx}
\ \\
\ \\
\begin{tabularx}{\textwidth}{ l l}
ID:& Block11\\
\hline
Text:& \parbox{10.3cm}{
The \$object is transferred by way of security to \$owner.Name.
}\\
\hline
Objects:& "owner:Person", "object:\$Object", "prop:PropertyRight"\\
\hline
Assignment:& \makecell[l]{
"owner.Name=Bank", 
"object=\$Block1\_shares",\\
"prop.Owner=\$owner",
"prop.Property=\$object"}
\end{tabularx}
}

\end{appendices}

\bibliography{arXiv}

\end{document}